\documentclass[10pt,journal,compsoc]{IEEEtran}

\usepackage{booktabs}
\usepackage{url}
\usepackage{amsmath}
\usepackage{latexsym} 
\usepackage{threeparttable}
\usepackage[ruled, vlined]{algorithm2e}
\usepackage{multirow}
\usepackage{multicol}
\usepackage{float}
\usepackage{subcaption}
\usepackage{enumitem}
\usepackage{circledsteps}
\usepackage{todonotes}
\usepackage{listings}
\usepackage{xcolor}
\usepackage{tikz}
\usepackage{colortbl}
\usepackage[T2A,T1]{fontenc}
\usepackage[utf8]{inputenc}
\usepackage[russian,english]{babel}

\usepackage{makecell}
\usepackage[misc,geometry]{ifsym}
\SetKwIF{If}{ElseIf}{Else}{if}{}{else if}{else}{end if}%

\definecolor{codegreen}{rgb}{0,0.5,0}
\definecolor{codegray}{rgb}{0.5,0.5,0.5}
\definecolor{codepurple}{rgb}{0.5,0,0.8}
\definecolor{backcolour}{gray}{0.97}
\lstdefinestyle{mystyle}{
    backgroundcolor=\color{backcolour},   
    commentstyle=\color{codegreen},
    keywordstyle=\color{magenta},
    numberstyle=\tiny\color{codegray},
    stringstyle=\color{codepurple},
    basicstyle=\ttfamily\footnotesize,
    breakatwhitespace=false,         
    breaklines=true,                 
    captionpos=b,                    
    keepspaces=true,                 
    numbers=left,                    
    numbersep=5pt,                  
    showspaces=false,                
    showstringspaces=false,
    showtabs=false,                  
    tabsize=2
}
\makeatletter
\newcommand{\removelatexerror}{\let\@latex@error\@gobble}
\makeatother

\newcolumntype{C}[1]{>{\centering\arraybackslash}p{#1}}
\newcommand*\ballnumber[1]{\tikz[baseline=(char.base)]{
            \node[shape=circle,draw,inner sep=0pt] (char) {#1};}}

\RestyleAlgo{boxruled}
\begin{document}
\title{A Side-channel Analysis of Sensor Multiplexing for Covert Channels and Application Profiling on Mobile Devices}
%\author{Anonymised for Review}
%\author{{\rm Carlton Shepherd$^{(\textrm{\Letter})}$, Jan Kalbantner, Benjamin Semal, and Konstantinos Markantonakis}\\Information Security Group\\Royal Holloway, University of London\\Egham, Surrey, United Kingdom\\{\rm \{carlton.shepherd, jan.kalbantner.2018, benjamin.semal.2018, k.markantonakis\}@rhul.ac.uk}}

\author{Carlton~Shepherd, Jan~Kalbantner, Benjamin~Semal, and Konstantinos~Markantonakis% <-this % stops a space
\IEEEcompsocitemizethanks{\IEEEcompsocthanksitem Carlton Shepherd is of Newcastle University, United Kingdom. The remaining authors are of the Information Security Group at Royal Holloway, University of London, Egham, Surrey, United Kingdom.\protect\\
E-mail: carlton.shepherd@ncl.ac.uk\\\newline
This work has been submitted to the IEEE for possible publication. Copyright may be transferred without notice, after which this version may no longer be accessible.}}% <-this % stops an unwanted space


\markboth{}%
{Shepherd \MakeLowercase{\textit{et al.}}}

%\renewcommand{\shortauthors}{Anonymised}



%\keywords{firmware, sensors, mobile security, Android}


\IEEEtitleabstractindextext{%
\begin{abstract}
Mobile devices often distribute measurements from physical sensors to multiple applications using software multiplexing. On Android devices, the highest requested sampling frequency is returned to all applications, even if others request measurements at lower frequencies.  In this paper, we comprehensively demonstrate that this design choice exposes practically exploitable side-channels using frequency-key shifting. By carefully modulating sensor sampling frequencies in software, we show how unprivileged malicious applications can construct reliable spectral covert channels that bypass existing security mechanisms. Additionally, we present a novel variant that allows an unprivileged malicious application to profile other active, sensor-enabled applications at a coarse-grained level. Both methods do not impose any special assumptions beyond accessing standard mobile services available to developers. As such, our work reports side-channel vulnerabilities that exploit subtle yet insecure design choices in Android sensor stacks.
\end{abstract}

}
\maketitle

\IEEEdisplaynontitleabstractindextext

% \fi
\IEEEpeerreviewmaketitle

\IEEEraisesectionheading{\section{Introduction}\label{sec:introduction}}

\IEEEPARstart{M}{obile} devices contain an array of sensors that measure the device's location, position, and ambient environment. The Android operating system currently supports over a dozen sensors, from accelerometers and gyroscopes to magnetic field, temperature, humidity and air pressure sensors~\cite{android:sensors_overview}. Modern devices host several multi-sensor integrated circuits (ICs) in millimeter-sized packages, driven by the proliferation of low-cost micro-electromechanical systems (MEMS).  Today, on-board sensors are used ubiquitously for implementing game controllers, detecting screen orientation changes, and performing gesture and human activity recognition~\cite{android:sensors_overview,gupta2016continuous,wang2012gesture,goel2012gripsense}. 

At run-time, multiple software applications may attempt to access measurements from one or more hardware sensors. A fitness tracker and a navigation app, for example, may both request measurements from a magnetometer sensor to detect the device's directional heading.  To resolve this, a widely used approach is for operating systems to use software multiplexing, where measurements are returned to \underline{\emph{all}} applications at the maximum requested sampling frequency. That is, if apps $A$ and $B$ request the same sensor at 100Hz and 50Hz respectively, then measurements are returned at 100Hz to both $A$ and $B$. Consequently, there is no guarantee that measurements will arrive at the specified rate if another app requests the \emph{same} sensor at a \emph{higher} frequency.

In this paper, we show how sensor multiplexing exposes practical software-controlled side-channels. In the first part of this work (\S\ref{sec:covert_channels}), we show how unprivileged applications can construct spectral covert channels for unauthorised inter-process communication (IPC). Briefly, this is achieved using a low-frequency carrier signal generated by the receiver application. A transmitting application can deterministically modulate the receiver's carrier signal by requesting the same sensor at a higher sampling frequency. Data is then transmitted using frequency-shift keying (FSK) between the transmission and carrier frequencies. For the first time, we present detailed experimental results showing that common mobile device sensors can be exploited on different devices from major manufacturers. Our approach enables the transmission of arbitrary bit-strings between two applications on the same device, including low resolution images. The method bypasses existing IPC protection mechanisms, e.g.\ application sandboxing, with low error depending on the chosen sensor and device.

In the second part of this paper (\S\ref{sec:application_fingerprinting}), we present a novel variant of this technique that enables coarse-grained application profiling.  The same approach of modulating a carrier frequency is used. However, this time, a malicious application establishes carrier signals for \emph{all} device sensors simultaneously. One of these signals is modulated when a sensor-enabled victim app starts requesting measurements at a higher frequency, which can be detected by the malicious app. We then perform a two-part study. Firstly, we show how various sensor sampling constants in the Android SDK~\cite{android:sensor_stack} can be detected with high accuracy and near real-time latency. These constants are designed to developers balance utility and battery consumption for various sensor use cases (e.g.\ games, detecting screen orientation changes, and UI interactions). In the second study, the top 250 Android applications according to AndroidRank~\cite{android:rank} (February 2022) are tested. We show how 1-in-5 of all tested applications (57/250; 22.8\%) used detectable sensor and sampling frequencies, which may be leveraged for user behavioural profiling. 

For both attacks, design and implementation information, detailed experimental results, and corrective recommendations are presented. Furthermore, we analyse how recent measures introduced in Android 9, intended to limit long-polling continuous sensors, are ultimately insufficient.  Our presented techniques do not impose any special requirements, e.g.\ rooting or kernel-mode access, beyond the use of sensor APIs available to developers through the Android Sensor SDK.

\subsection{Contributions}
We extensively examine the extent to which sensor measurement multiplexing can be used for: \ballnumber{1} spectral covert channels for app-wise IPC; and \ballnumber{2} profiling sensor-enabled victim applications. The methods are evaluated against a large range of standard mobile sensors using devices from separate major OEMs, thus affecting millions of devices globally. Moreover, we show how 22.8\% of the top 250 Android applications are vulnerable to our profiling technique to some degree. Proof-of-concept code for the attacks is made open-source to facilitate future research.\footnote{Proof-of-concept repository URL:~\url{https://github.com/cgshep/android-multiplexing-security-pocs}}

\subsection{Responsible Disclosure}
The results in this paper were disclosed to Google's Android Security Team on 3rd June 2021 under a 90-day disclosure period. They were acknowledged on the same day, a severity assessment was provided on the 21st June 2021. The issues will be addressed in a forthcoming major Android release. A bug bounty was awarded to the authors, which they donated to a charity of Google's choosing.

\section{Related Work}
\label{sec:related_work}

Current research has focussed on timing side-channels using CPU cache contents  (\textsc{Prime+Probe} \cite{osvik2006cache}, \textsc{Flush+Reload}~\cite{yarom2014flush}, \textsc{Flush+Flush}~\cite{gruss2016flush}, and their variants, e.g.\ \cite{lipp2016armageddon,cho2018prime,maurice2015c5,liu2015last}), exploiting contention in DRAM memory controllers~\cite{semal2020leaky,semal2020one} and GPU access patterns~\cite{naghibijouybari2018rendered,jiang2016complete}.  Additionally, mobile sensors have found wide utility for out-of-band channels using the device's ambient environment. MEMS gyroscopes have been used as low-frequency microphones for detecting ambient speech~\cite{michalevsky2014gyrophone} and accelerometers for inferring touch interactions~\cite{xu2012taplogger,owusu2012accessory,cai2011touchlogger}.  CPU core temperature sensors have been used for same-core and cross-core thermal side-channels, where victim processes deterministically increase the core temperature observed by a malicious application~\cite{masti2015thermal,bartolini2016capacity,long2018improving}. Novak et al.~\cite{novak2015physical} described how camera flashes, vibrations, and device speakers can be used to transmit data to a (same-device) receiver. Block et al.~\cite{block2017autonomic} used a similar approach with ultrasonic frequencies. In 2016, Matyunin et al.~\cite{matyunin2016covert} showed how EM emissions of I/O operations can be detected by mobile magnetic field sensors. In 2019, the MagneticSpy~\cite{matyunin2019magneticspy} attack showed how a malicious app could profile same-device apps and web pages using the magnetic field sensor with $>$90\% accuracy.  

To distinguish this work, we do not use physical transmission media. Instead, we exploit side-effects of the \emph{software interfaces} used for resolving simultaneous accesses to sensing hardware. We take inspiration from Ulz et al.~\cite{ulz2019sensing}, who presented methods for creating covert channels on embedded systems using: \ballnumber{1}, unused/reserved registers of sensing microcontroller units (MCUs); \ballnumber{2}, reading/writing to MCU register configuration bits; and \ballnumber{3}, timing differences between the activation of sensor MCUs. The authors posit that observable effects of different requested sampling periods could be used for frequency-based information encoding. However, detailed experimental evidence was not presented using various sensors and devices, and their error rates.  In this work, we comprehensively and empirically show that continuous sensors on Android devices are vulnerable to covert channels using sensor measurement multiplexing. Moreover, we develop a novel attack variant for profiling sensor-enabled victim applications.

\section{The Android Sensor Stack}
\label{sec:sensor_stack_internals}

The Android framework supports 13 hardware- and software-based sensors.\footnote{We exploit continuous sensors defined in the Android Sensor SDK~\cite{android:sensors_overview}. GPS location, sound levels, and detecting nearby Bluetooth and Wi-Fi devices have also been treated as so-called `sensors'~\cite{gurulian2017effectiveness,halevi2012secure,novak2015physical,block2017autonomic}. However, these use modalities do not use sensor multiplexing and are not vulnerable to the same attacks.}   Hardware sensors take measurements directly from a sensor integrated circuit (IC), e.g.\ a MEMS gyroscope. Software (virtual) sensors derive their measurements using signals from one or more hardware sensors within a sensor hub or the operating system. The exact separation between physical and virtual sensors is OEM-dependent. As a guide, accelerometers, gyroscopes and magnetic field sensors are usually implemented in hardware, while rotation vector and linear acceleration sensors are common virtual sensors~\cite{android:sensors_overview}. 

\begin{figure}
    \centering
    \includegraphics[width=0.97\linewidth]{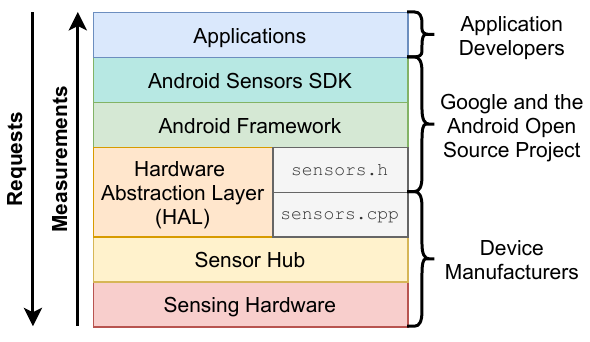}
    \caption{Components, corresponding owners, and information flow of the Android sensor stack (adapted from \cite{android:sensor_stack}).}
    \label{fig:sensor_stack}
\end{figure}

 The Android sensor stack (Figure~\ref{fig:sensor_stack}) comprises several layers for translating raw physical values, which are device- and OEM-dependent, into standard interfaces that can be used by application developers. At an application level, developers use the \texttt{SensorManager} class for sensor enumeration and for measurement acquisition. Measurements are requested using the \texttt{SensorEventListener} interface and by calling the \texttt{registerListener} with the desired sensor and sampling rate. The SDK routes requests to the HAL---a single-client layer that abstracts OEM-specific firmware---which is shared by Android OS and all user applications. The HAL communicates directly with sensing hardware or, optionally, to a sensor hub for pre-processing measurements without waking the main application processor.  The HAL returns the resulting data to the Android framework using FIFO fast message queues (FMQs), thus avoiding kernel involvement~\cite{android:fmqs}. The framework returns sensor data to the requesting application as \texttt{SensorEvent} objects on an event-driven and first-in first-out (FIFO) basis to its \texttt{SensorEventListener} interface. When multiple apps register requests to a single sensor simultaneously, the multiplexing mechanism is used irrespective of whether physical or virtual sensors are requested.

\section{Building Covert Channels from Sensor Multiplexing}
\label{sec:covert_channels}

We recognised that a lack of appropriate controls at the Sensor API layer enables applications to influence the sampling frequencies of other applications. This section develops an FSK-based covert channel based on this observation.  

\subsection{Threat Model}

The covert channel involves two colluding applications---a transmitter, $Trn$, and receiver, $Recv$---who wish to establish a uni-directional communication channel that bypasses Android's IPC security mechanisms. The attack scenario assumes that $Trn$ has access to security- or privacy-sensitive data; for example, permission to access SMS data or GPS co-ordinates. $Trn$ wishes to send this data to $Recv$, which does not possess such permissions. However, $Recv$ \emph{does} have the ability to extract data from the device, such as permissions for accessing the internet to leak data to a remote server. Crucially, the user may not wish to download $Recv$ if it requests permissions to sensitive data \emph{and} a data transmission medium. Our proposed channel enables $Trn$ to leak data to $Recv$ using only standard methods for receiving sensor measurements.  These are implemented as \emph{separate} services launched independently by $Trn$ and $Recv$, which may be disguised as two legitimate applications.\footnote{One-to-one communication is described for simplicity. However, our approach can be used without alteration as a covert broadcast channel to multiple receivers. This is possible because sensor multiplexing modulates the sampling frequency of \emph{all} applications.}

\subsection{Channel Design}
\label{sec:channel_design}

The channel relies on $Trn$ and $Recv$ targeting a shared sensor, $S$. Initially, $Recv$ registers a sensor listener for $S$ using a long (slow) sampling period, $T_{c}$. After this, $Trn$ repeatedly registers and unregisters listeners for $S$ using a faster sampling period, $T_{tr}$. The repeated registering/un-registering corresponds to the bits of information that $Trn$ it wishes to transmit. Due to the multiplexing phenomenon, $Recv$'s \emph{observed} sampling period will modulate between $T_c$ and $T_{tr}$.  After each listener (un-)registration, $Trn$ must wait for a short time period (pulse width), $w$, to allow the new sampling period to be multiplexed into $Recv$'s signal. Note that the channel is not limited to two frequencies, $T_{c}$ and $T_{tr}$. It is possible to register multiple sampling periods at higher frequencies than $T_{tr}$, which are modulated at their own rates. Using this observation, we constructed a transmission protocol using four sampling periods: \ballnumber{1} to define the start of a message transmission, i.e.\ a syncword, $T_{sync}$; \ballnumber{2} to indicate transmission endings, $T_{end}$; \ballnumber{3} $Recv$'s carrier period, $T_{c}$; and \ballnumber{4} $Trn$'s modulating period, $T_{tr}$. By setting appropriate sampling periods and pulse widths---explored in \S\ref{sec:cc_implementation}---an FSK-based binary transmission channel is established.

The channel involves three stages:

\begin{figure}
    \centering
    \includegraphics[width=\linewidth]{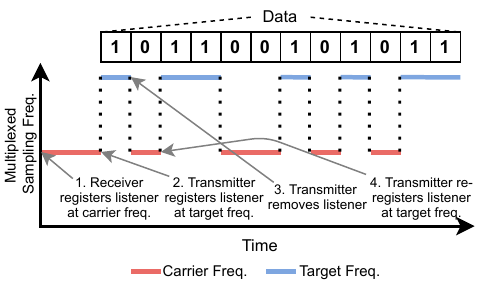}
    \caption{Simplified bit transmission procedure; the multiplexed frequency is observed by both applications.}
    \label{fig:highlevel}
\end{figure}

\textbf{1. Channel initialisation}: $Recv$ registers measurements using a target sensor, $S$, with period $T_{c}$. $Trn$ registers a listener for the \underline{\emph{same}} sensor, $S$, but at a significantly higher frequency, $T_{sync}$, to signal the start of a transmission. 

\textbf{2. Data transmission} (Figure~\ref{fig:highlevel}): $Trn$ unregisters the listener with sampling period $T_{sync}$. $Trn$ transmits a bit-encoded secret, $v = [b_0, b_1, \dots, b_n]$, by registering a new listener at $T_{tr}$ when $b_i = 1$ for time $w$.  Conversely, $Trn$ unregisters the listener and waits for time $w$ to transmit $b_i = 0$.  This causes $Recv$'s sampling frequency to return to $T_{c}$. In the background, $Recv$ observes these frequency changes and appends the bit values to an internal buffer. 

\textbf{3. Channel termination}: $Trn$ ends the transmission by registering a listener with period $T_{end}$. After detecting this, $Recv$ conducts any post-transmission operations, such as sending the buffer to an external server.

\subsection{Implementation}
\label{sec:cc_implementation}

State machines were implemented for managing our protocol using two separate Android applications for $Trn$ and $Recv$. The full implementations are released as open-source software (see \S\ref{sec:introduction}). The applications targeted \emph{continuous} sensors on the devices under test, which deliver measurements as \texttt{SensorEvent} objects as soon as they are available from the HAL. Continuous reporting sensors comprise the majority of sensors available in Android~\cite{android:sensors_overview}.  In this work, the accelerometer (AC), gyroscope (GY), gravity (GR), linear acceleration (LA), magnetic field (MF), and rotation vector (RV) sensors were successfully utilised. While theoretically vulnerable, sensors with other reporting modes, e.g.\ one-shot and on-change reporting, cannot be leveraged as effectively. These modalities require constant user interaction for consistent event generation.

\subsubsection{Test Devices and Hardware Information}

\begin{table*}
\centering
\caption{Sensing hardware and their minimum supported sampling periods ($\mu$s).}
\label{tab:min_sampling_frequency}
\begin{threeparttable}
\begin{tabular}{@{}r|cc|cc|cc@{}}
\toprule
\textbf{} & \multicolumn{2}{c|}{\textbf{Xiaomi Poco F1}} & \multicolumn{2}{c|}{\textbf{Motorola Moto G5}} & \multicolumn{2}{c}{\textbf{Google Pixel 4A}} \\
\textbf{Sensor} & \textbf{Vendor} & \textbf{Min. Period} & \textbf{Vendor} & \textbf{Min. Period} & \textbf{Vendor} & \textbf{Min. Period} \\ \midrule
AC & Bosch BM160 & 2500 &  Bosch*  & 10000 & STM LSM6DSR & 2404 \\
GR & Qualcomm* & 5000 & Motorola* & 10000 & Google* & 5000 \\
GY & Bosch BM160 & 5000 & Bosch* & 5000 & STM LSM6DSR & 2404  \\
LA & Qualcomm* & 5000 & Motorola & 10000 & Google*  & 20000   \\
MF & AKM AK0991x & 10000 & --- & --- & STM LIS2MDL & 10000    \\
RV & Xiaomi* & 5000 & Bosch* & 5000  & Google*  & 5000   \\ \bottomrule
\end{tabular}
\begin{tablenotes}
\item *: Specific model not known, ---: Not supported.
\end{tablenotes}
\end{threeparttable}
\end{table*}

Three test devices from separate OEMs were evaluated:
\begin{enumerate}
    \item \emph{Xiaomi Poco F1} with a Qualcomm Snapdragon 845 (octo-core at 2.8GHz, 6GB RAM) and Android 10 (build QKQ1.190828.002). Cost: \textsterling 309/$\sim$\$370 USD (2019).
    \item \emph{Google Pixel 4A} with a Qualcomm SDM730 Snapdragon 730G (octo-core at 2x2.2GHz and 6x1.8GHz, 6GB RAM) and Android 11 (build RQ2A.210405.005). Cost: \textsterling 349/$\sim$\$450 USD (2020).
    \item \emph{Motorola Moto G5} with a Qualcomm MSM8937 Snapdragon 430 (octo-core at 1.2GHz, 2GB RAM) and Android 8.1 (build OPP28.82-19-4-2). Cost: \textsterling 120/$\sim$\$170 USD (2017).
\end{enumerate}

The covert channel leverages changes in sampling frequencies, thus the fastest supported sampling rate acts as a hard throughput limit for a given sensor.  Sensing hardware information can be gathered using the \texttt{List<Sensor> getSensorList (int type)} function from the \texttt{SensorManager} class, where \texttt{int type} is \texttt{Sensor.TYPE\_ALL}. Next, \texttt{getMinDelay()} can be used to find the minimum latency allowed between two \texttt{SensorEvent} objects in microseconds for continuous sensors. We used this approach to find the minimum supported sampling periods for each device and sensor, including their hardware models (where known). This is shown in Table~\ref{tab:min_sampling_frequency}.

\subsubsection{Challenges}
\label{sec:param_setting}

During preliminary experiments, it became evident that the protocol implementation must overcome two obstacles:

\begin{enumerate}[start=1,align=left,labelwidth=1ex,label={[O\arabic*]:}]
\item Measurements from mobile device sensors are subject to significant temporal jitter.
\end{enumerate}

\begin{figure*}[t!]
    \centering
    \includegraphics[width=0.96\linewidth]{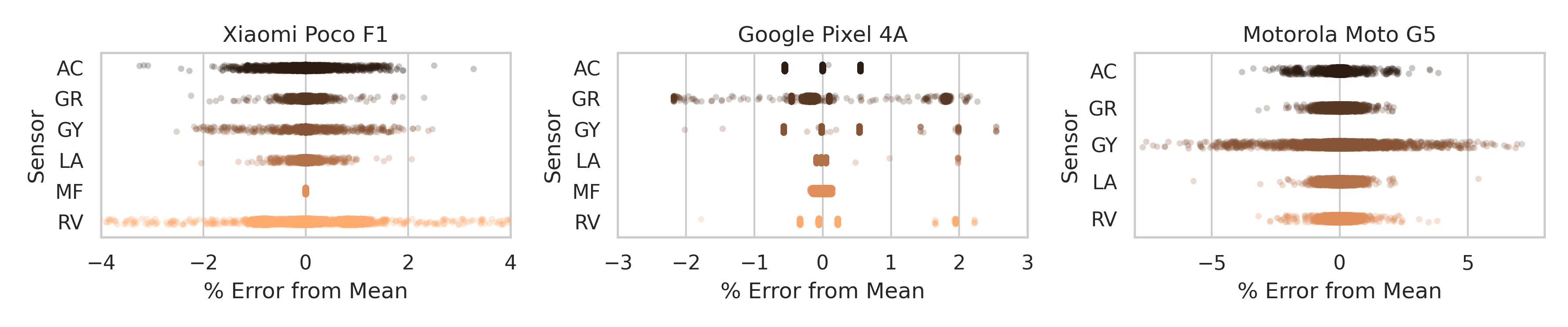}
    \caption{Inferred sampling period distributions using the lowest supported periods for each device and sensor.}
    \label{fig:jitter}
\end{figure*}
\begin{figure*}[t!]
    \centering
    \includegraphics[width=\linewidth]{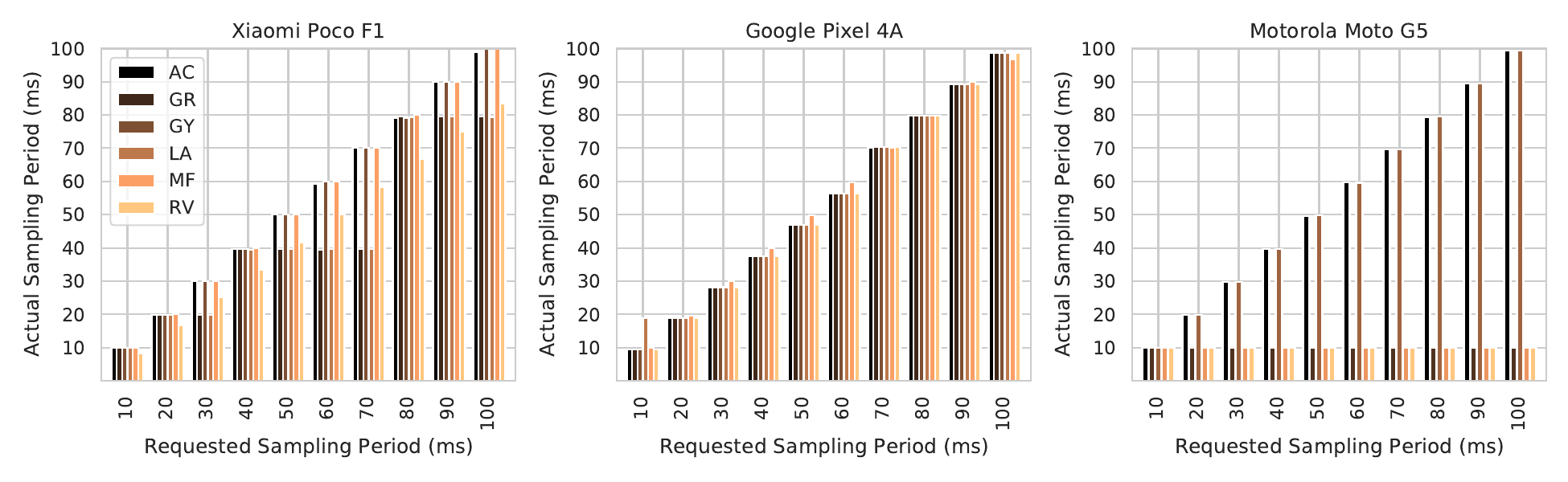}
    \caption{Sensor measurement sampling responses.}
    \label{fig:sampling}
\end{figure*}

Sampling jitter complicated the ability to precisely infer the current sampling period. According to the sensor batching documentation:~\emph{``Physical sensors sometimes have limitations on the rates at which they can run and the accuracy of their clocks. To account for this, the actual sampling frequency can differ from the requested frequency as long as it satisfies the [following] requirements.''}~\cite{android:batching}. If the requested frequency is below the min. frequency, then between 90\%--110\% of the min. frequency must be returned. If the request is between the min. and max. frequency, then 90\%--220\% of the requested frequency must be returned. Lastly, if the request is above the max. supported frequency, then 90\%--110\% of the max. frequency must be returned, but below 1100Hz~\cite{android:batching}. Measurement jitter was studied by sampling each sensor at its maximum supported rate on each device (Table \ref{tab:min_sampling_frequency}). 10,000 measurements were collected per sensor and per device, before calculating the inferred sampling period using the microsecond-level time between consecutive \texttt{SensorEvent} objects. (Sensor measurements are managed internally using FIFO FMQs, thus accurate period inferencing is possible using the timestamps of sequential \texttt{SensorEvents}~\cite{android:hal2}).

Figure~\ref{fig:jitter} shows the distributions of the inferred sampling periods.  The disparity in device- and sensor-wise jitter likely arises from varying sensor manufacturing tolerances. This is compounded by implementation differences in proprietary, OEM-specific sensor HALs. To overcome this, an error threshold, $\epsilon = 0.1$ (10\%), was used as a detection tolerance for each protocol frequency band. In general, the MF sensor exhibited the lowest jitter ($<$0.5\% error from the mean, Xiaomi Poco F1 and Google Pixel 4A; no MF sensor on the Moto G5). The AC, GR and LA sensors were the next best performing ($<$1--2\%, Xiaomi Poco F1; $<$2\%, Pixel 4A; $<$2.5\%, Moto G5). The GY and RV exhibited the largest jitter ($<$2--4\%, Xiaomi Poco F1; $<$2\%, Pixel 4A; $<$2.5--5\%, Moto G5).  It is noteworthy that jitter tends to increase when the minimum supported sampling period decreases.

\begin{enumerate}[start=2,align=left,labelwidth=1ex,label={[O\arabic*]:}]
\item The \emph{requested} sampling period can significantly differ from the \emph{actual} period at which measurements are returned.
\end{enumerate}

On the Xiaomi Poco F1, the GR, LA and RV sensors exhibited step function-like behaviour, and regularly over-sampled the requested period (Figure~\ref{fig:sampling}). In the worst case, the GR and LA sensors sampled at a $\sim$40\% faster rate than requested.  This differed between IC manufacturers on the same device: the least accurate (GR and LA) belonged to an undisclosed Qualcomm IC, while the most accurate---the AC, GY and MF sensors---were provided by a Bosch BM160 and AKM AK0991x. The RV sensor also consistently over-sampled, but without the step-like behaviour of the GR and LA sensors.  Generally, the Google Pixel 4A exhibited the best response rates: all actual periods were within $>$95\% of the requested period, exceeding 99\% in most cases. Only two Moto G5 sensors responded well: the AC and GY sensors (from a Bosch IC). The remaining Moto G5 sensors returned only a single frequency irrespective of the requested frequency. This prevented the use of multiple frequencies for our channel, so these sensors were removed from further consideration. Note that, following tests with another Pixel 4A and Poco F1, we can confirm that the same parameters can be used unchanged between device models.

\subsection{Evaluation}
\label{sec:eval}

This section evaluates the developed covert channel using the six continuous sensors from \S\ref{sec:cc_implementation}.

\subsubsection{Methodology}
\label{sec:covert-methodology}
The average error rate and throughput (bit rate) of the proposed covert channel was evaluated for each device. The pulse width---the length of time between transmitting individual bits---was the dominating factor for the channel's throughput. Intuitively, the shorter the pulse width, then the greater the bit rate.
Protocol sampling bands were based on the maximum supported frequency as the base signal.  Other frequencies were derived from multiples of this period according to the relationship in Equation~\ref{eq:frequency_hierarchy}. Where this method could not be used, the next distinct sampling rate at a longer period was selected. The full set of sampling bands used on each device is provided in Appendix~\ref{sec:appendix_params}.

\begin{equation}
    Trn(T_{end}) \ll Trn(T_{sync}) \ll Trn(T_{tr}) \ll Recv(T_{c})
    \label{eq:frequency_hierarchy}
\end{equation}

We measured 100 transmissions of 64--256 bit random sequences on a per-sensor and per-device basis. For each configuration, pulse widths were evaluated in the following range: the minimum value was the shortest period before total channel failure was observed (no bits detected by $Recv$). This value was increased until error-free transmission was achieved (across 100 bit-strings) or the error rate was unchanged.  In total, 74 configurations were evaluated across the devices and sensors.

Data files were created on $Trn$ and $Recv$ at each channel instance, containing the transmitted and received bit-strings, and timestamps for each sensor and pulse width. The timestamps corresponded to immediately before the syncword transmission ($Trn$-side) and after the post-amble signal ($Recv$-side). The data files from the experiments---completed over two days by three volunteers---were retrieved from the devices for analysis. The average (median) bit rate was then calculated for each pulse width, device, and sensor. The transmission error rate was measured using the Levenshtein (edit) distance between the transmitted and received bit-strings.  This metric, also used in \cite{maurice2015c5,cabuk2004ip,krosche2018dpid,lin2015designing,liu2015last}, counts the insertions, deletions and substitutions for reverting the received bit-string to the transmitted one. Bit deletions/omissions were the primary source of transmission errors. This is due to Android's lack of real-time measurement delivery guarantees, which became increasingly pronounced at higher sampling frequencies.

\subsubsection{Results}
\label{sec:covert_channel_results}

\begin{figure*}
    \centering
    \includegraphics[width=0.95\linewidth]{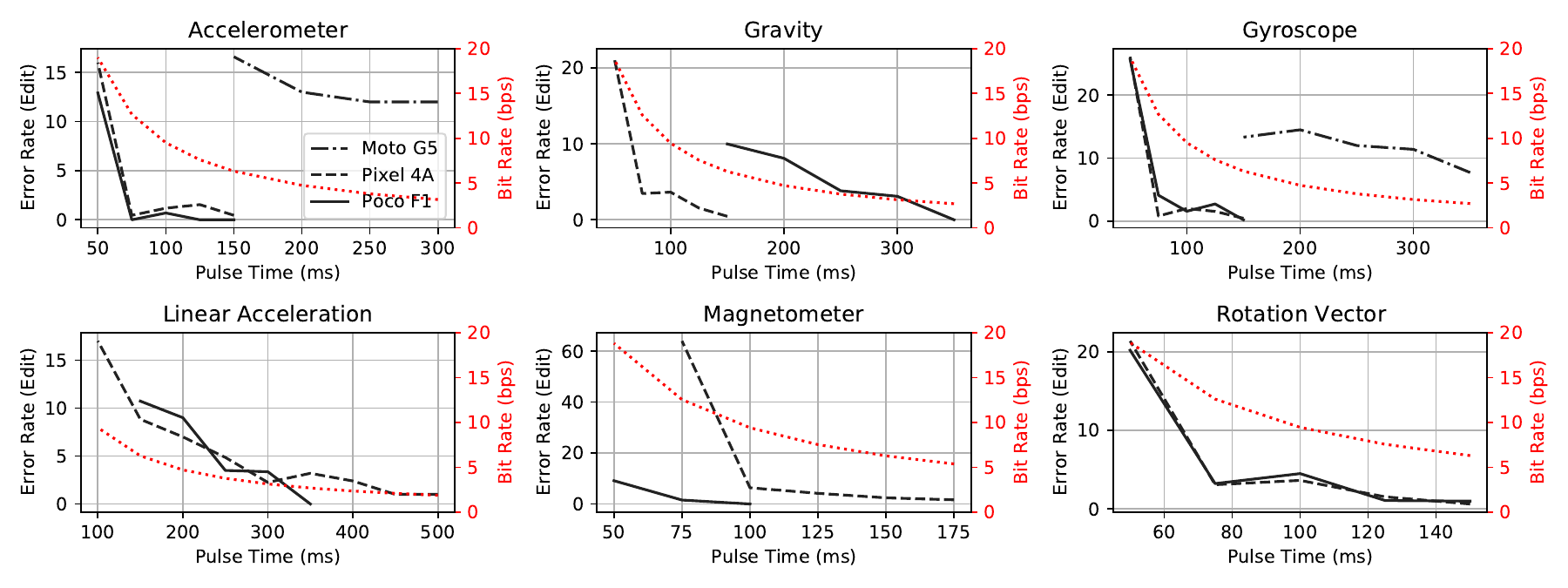}
    \caption{Average channel error rates and bit rates for each device and sensor.}
    \label{fig:error_results}
\end{figure*}
\begin{figure*}
    \centering\includegraphics[width=0.91\linewidth]{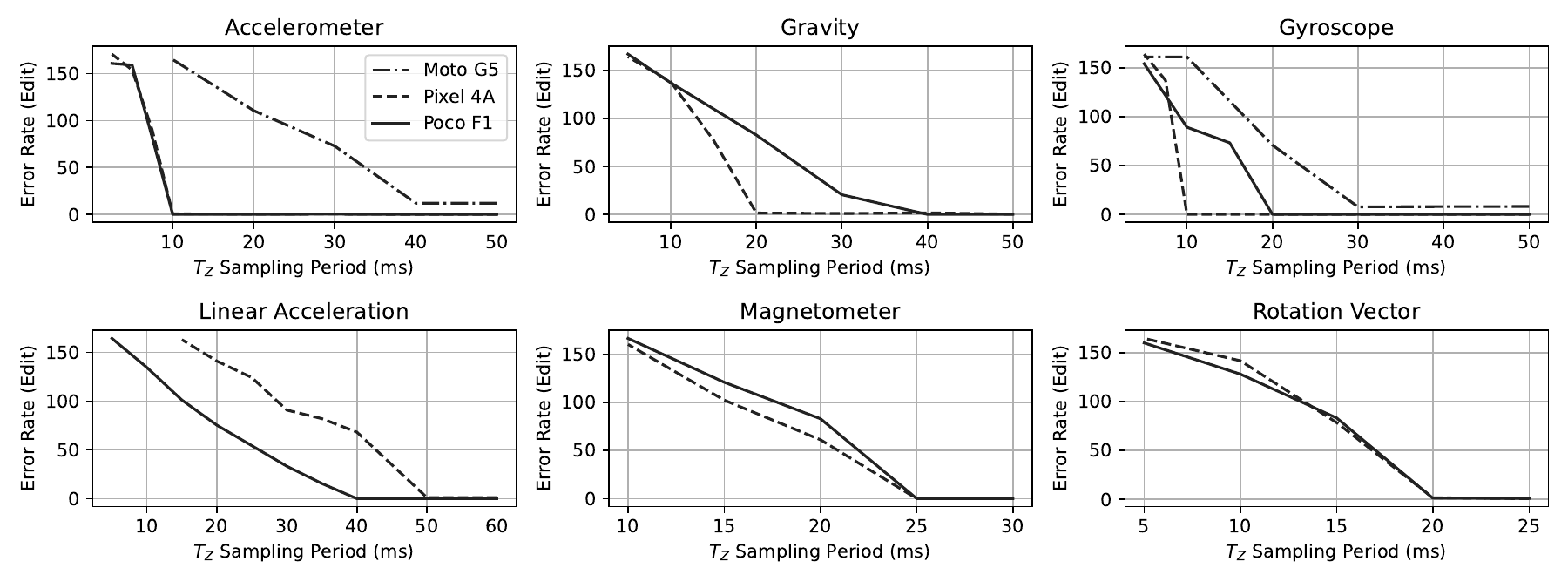}
    \caption{Average channel error rates under interference from a third application using a sampling period of $T_Z$.}
    \label{fig:interference}
\end{figure*}

The experimental results are presented in Figure~\ref{fig:error_results}. For each sensor, very low transmission error rates were attained on at least one evaluation device. The best-case rates are shown in Table~\ref{tab:best_case_errors}. The AC sensor was able to achieve near-zero error rate using a pulse width of 75ms (Pixel 4A and Poco F1), corresponding to a $\sim$10bps bit-rate. Both devices produced zero error using a 150ms width with a bit-rate at 5.1bps. The Moto G5 AC sensor, however, produced significant error ($\sim$12 edit distance), even at long widths (e.g.\ 300ms). MF was the next best performing sensor, achieving error-free transmission for the Poco F1 (100ms, 9.62bps). However, the Pixel 4A required a longer width to achieve the same results (175ms, 5.08bps). The Pixel 4A's GR sensor achieved near error-free transmission (0.215 average edit distance) with a 7.28bps bit rate using a 150ms width. The Poco F1 did not achieve this until a 350ms pulse width (2.81bps). In the best case, the RV sensor on the Pixel 4A achieved the lowest error rate of a 0.636 edit distance at 15ms (6.70bps). The LA sensor fared significantly worse, error-free transmission at 2.89bps was still achieved with the Pixel 4A.

\begin{table}
\caption{Best case error rates (average of 100 iterations).}
\label{tab:best_case_errors}
\centering 
\resizebox{\columnwidth}{!}{%
\begin{tabular}{@{}r|c|c|c|c@{}}
\toprule
\textbf{Sensor} & \textbf{Pulse (ms)} & \textbf{Error (Edit)} & \textbf{Bit Rate (bps)} & \textbf{Device} \\ \midrule
\textbf{AC} & 150 & 0 & 5.10 & Poco F1 + Pixel 4A \\
\textbf{GR} & 150 & 0 & 7.37 & Pixel 4A \\
\textbf{GY} & 150 & 0.215 & 7.28 & Poco F1  \\
\textbf{LA} & 350 & 0 & 2.89 & Poco F1  \\
\textbf{MF} & 100 & 0 & 9.62 & Poco F1  \\
\textbf{RV} & 150 & 0.636 & 6.70 & Pixel 4A  \\ \bottomrule
\end{tabular}
}
\end{table}

\begin{figure}[t!]
\centering
\begin{subfigure}{.23\linewidth}
  \centering
  \includegraphics[width=\linewidth]{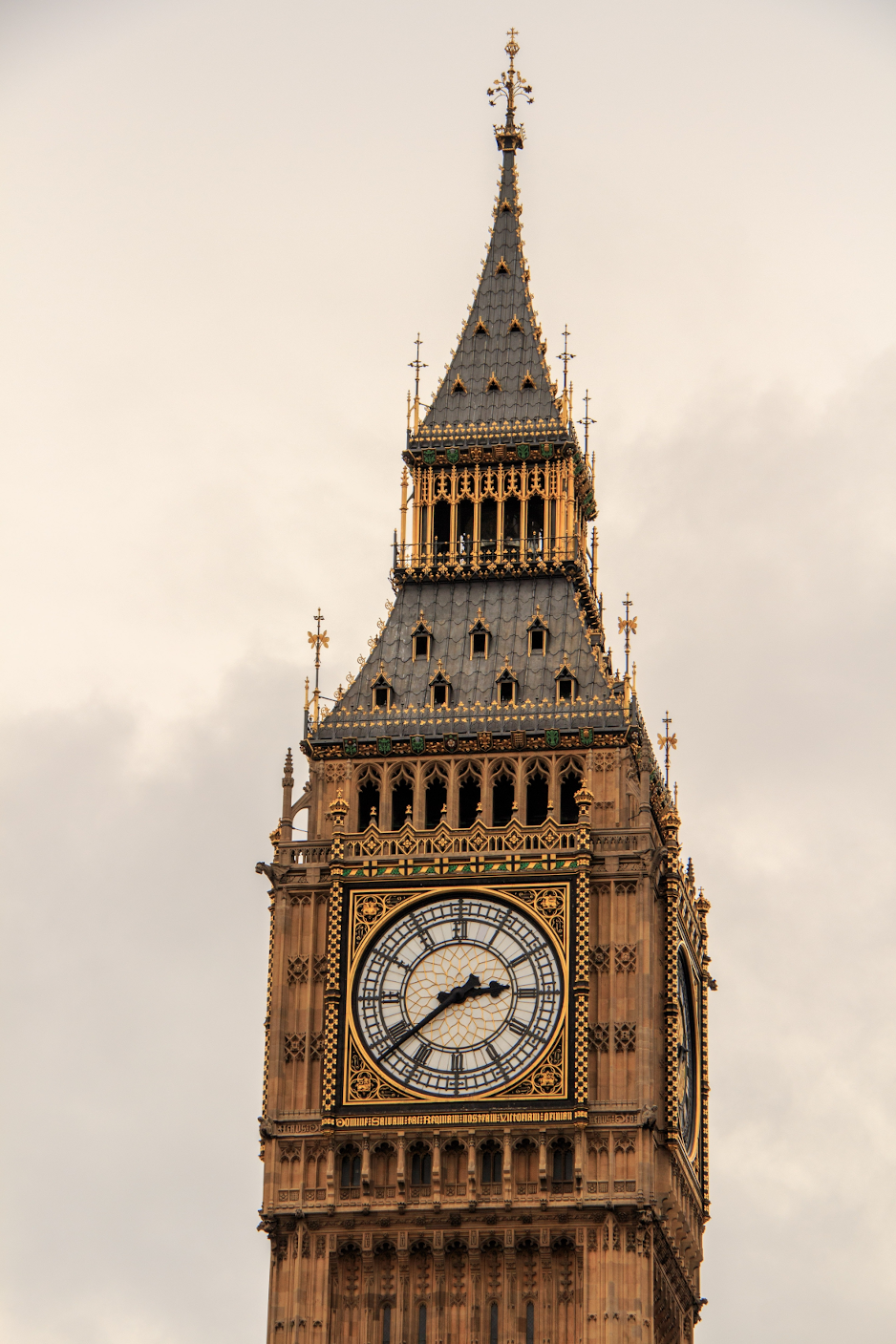}
  \label{fig:sfig1}
\end{subfigure}%
\begin{subfigure}{.23\linewidth}
  \centering
  \includegraphics[width=\linewidth]{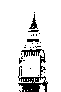}
  \label{fig:sfig2}
\end{subfigure}
\begin{subfigure}{.23\linewidth}
  \centering
  \includegraphics[width=\linewidth]{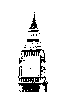}
  \label{fig:sfig3}
\end{subfigure}
\begin{subfigure}{.23\linewidth}
  \centering
  \includegraphics[width=\linewidth]{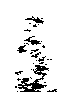}
  \label{fig:sfig4}
\end{subfigure}
\begin{subfigure}{.23\linewidth}
  \centering
  \includegraphics[width=\linewidth]{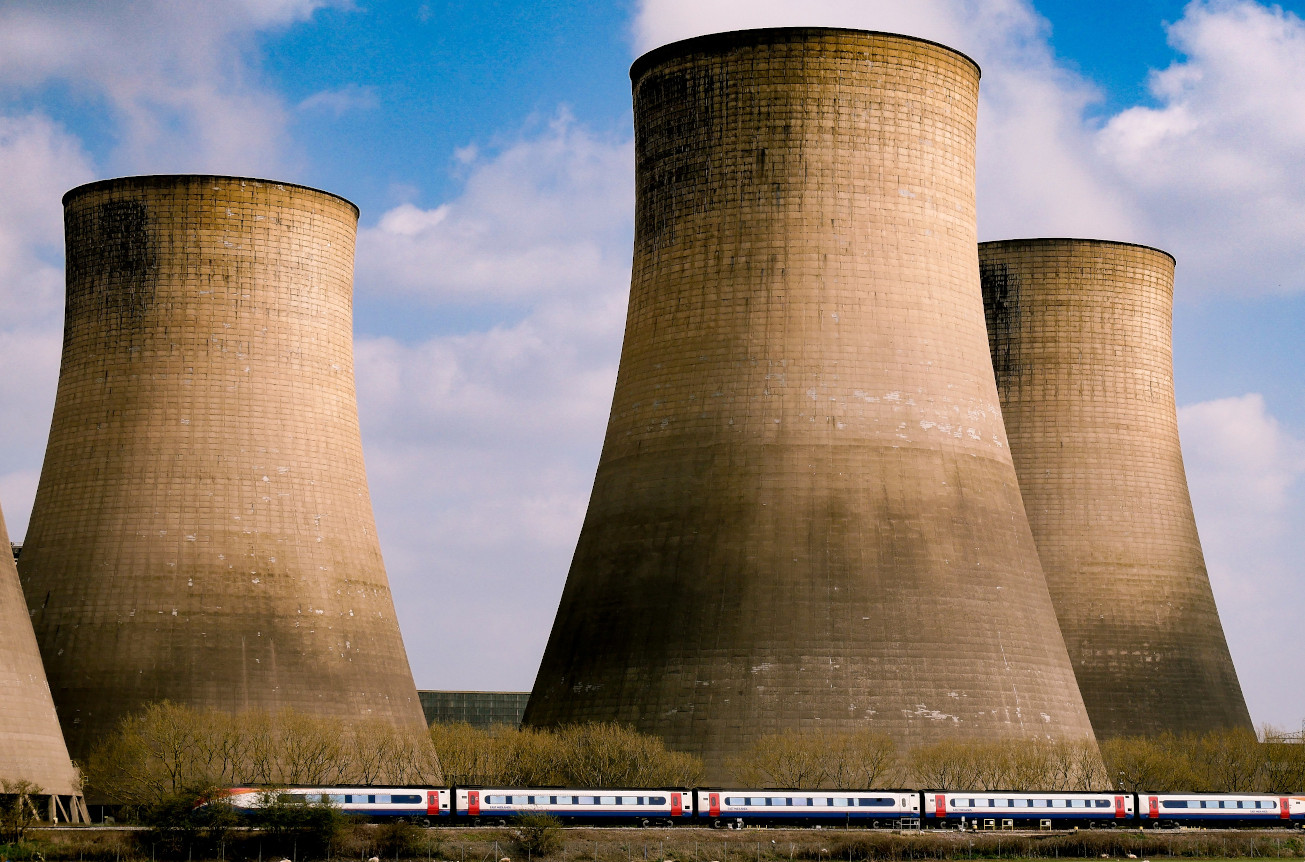}
  \label{fig:sfig2}
\end{subfigure}
\begin{subfigure}{.23\linewidth}
  \centering
  \includegraphics[width=\linewidth]{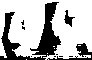}
  \label{fig:sfig2}
\end{subfigure}
\begin{subfigure}{.23\linewidth}
  \centering
  \includegraphics[width=\linewidth]{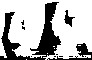}
  \label{fig:sfig2}
\end{subfigure}
\begin{subfigure}{.23\linewidth}
  \centering
  \includegraphics[width=\linewidth]{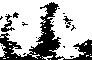}
  \label{fig:sfig2}
\end{subfigure}
\begin{subfigure}{.23\linewidth}
  \centering
  \includegraphics[width=\linewidth]{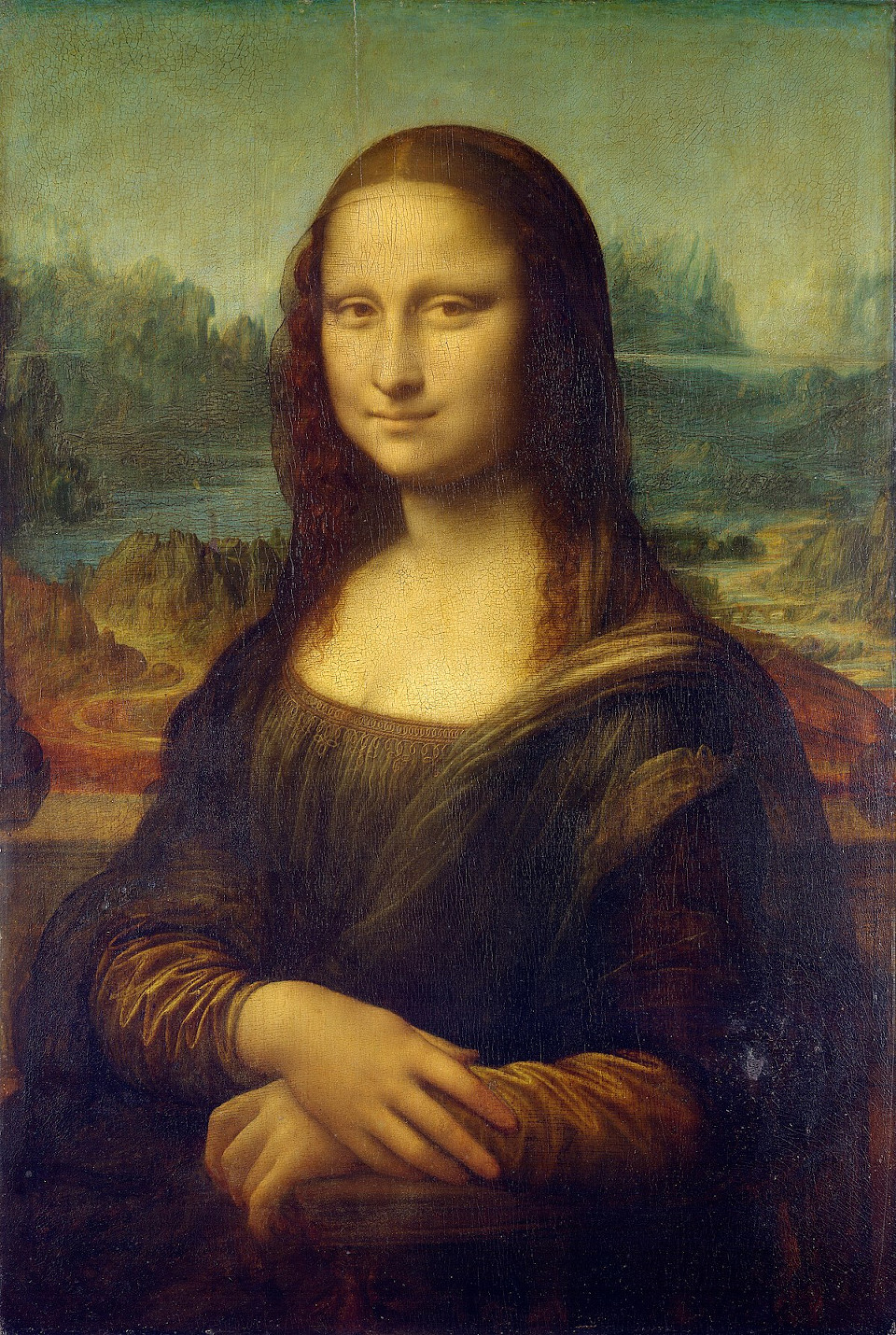}
  \caption{Original.}
  \label{fig:sfig2}
\end{subfigure}
\begin{subfigure}{.23\linewidth}
  \centering
  \includegraphics[width=\linewidth]{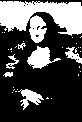}
  \caption{Encoded.}
  \label{fig:sfig2}
\end{subfigure}
\begin{subfigure}{.23\linewidth}
  \centering
  \includegraphics[width=\linewidth]{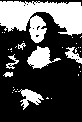}
  \caption{Clean.}
  \label{fig:sfig2}
\end{subfigure}
\begin{subfigure}{.23\linewidth}
  \centering
  \includegraphics[width=\linewidth]{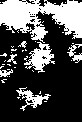}
  \caption{Noisy.}
  \label{fig:sfig2}
\end{subfigure}
\caption{Example image transmissions. Figures (a) and (b) are sent by $Trn$; (c) and (d) are seen by $Recv$. \emph{(Device: Xiaomi Poco F1; clean channel: MF at 150ms; noisy: GR at 250ms.)}}
\label{fig:channel_imgs}
\vspace{-0.5cm}
\end{figure}

Interestingly, the channel performed most effectively on the highest cost device (Pixel 4A). Conversely, the oldest, lowest cost handset (Moto G5) exhibited high error and had only two exploitable sensors. Error-free channels were not observed on this device. This leads to the conjecture that handsets manufactured with higher quality component, are particularly susceptible to low error, high bit-rate covert channel transmissions. Beyond random bit-strings, we also investigated transferring images and numeric values. This better represents a realistic scenario where a disguised legitimate app leaks sensitive data to a receiver. Accordingly, we experimented with transferring photographs using a three-stage process: \ballnumber{1} encoding the desired image using a one-bit colour palette; \ballnumber{2} transmitting the encoded bit-string; and \ballnumber{3} exporting the bit-string in a suitable image format, e.g.\ PNG. Stages \ballnumber{1} and \ballnumber{3} utilised a Python script using NumPy, SciPy, and the Python Image Library (PIL). Example image transfers are shown in Figure~\ref{fig:channel_imgs}. We were also successful in transmitting GPS coordinates by encoding and segmenting decimal values as 4-bit strings. Communicating natural language is more difficult, e.g.\ leaking SMS records and instant messages. This requires a character encoding scheme that retains reasonable throughput. Given the low bit rate---$<$10bps for near error-free transfers---transmitting individual characters take $\sim$1000ms using UTF-8, which limits its feasibility to short strings.

\subsubsection{Error Rates Under Interference}
\label{sec:error-rates-interference}

We have seen how $Trn$ and $Recv$ can establish frequency bands without interference. However, it is conceivable that the user may launch (or has already launched) a third sensor-enabled application, $Z$. This app may use measurements from the same shared sensor as $Trn$ and $Recv$. Interference can occur depending on $Z$'s requested sampling period, $T_Z$, according to the channel design rules in \S\ref{sec:channel_design}. 

If $T_Z > T_{c}$ (lower frequency), then the covert channel will not be affected because $T_{c}$ will be used as the dominant frequency for all apps ($Trn$, $Recv$, and $Z$). However, if $T_Z \approx T_{tr}$ then all channel data will be interpreted as binary `1's, as $T_Z$ will always override FSK transitions to $T_{c}$ (which denotes binary `0's). If $T_Z$ is between $T_c$ and $T_{tr}$, then the interpreted value will vary depending on $T_Z$'s closeness to either period. 

To analyse this effect, we began with the best case sampling bands from the previous section (see Tables~\ref{tab:pixel4a_cc_periods}--\ref{tab:motog5_cc_periods} in Appendix~\ref{sec:appendix_params}). We then transmitted 100 random 64--256 bit binary strings between $Trn$ and $Recv$ using the same method as in \S\ref{sec:covert-methodology}. However, this time prior to each transmission, we opened a third app beforehand that acted as $Z$ (repurposed from the code-base of $Recv$). $Z$ creates a sensor listener for a given sensor and sampling period, $T_Z$. The value of $T_Z$ was changed through $T_c$ to the shortest (fastest) frequency, $T_{end}$ (where channel failure is expected). The average error rate was computed at each value, and repeated for each applicable sensor and device.

The results are given in Figure~\ref{fig:interference}. Notably, we observe that the error rate increases dramatically as the sampling period of the third application, $T_Z$, is decreased through to the lowest period. Due to the multiplexing phenomenon, $T_Z$ is received by $Recv$, overriding any slower frequencies in the hierarchy (Equation~\ref{eq:frequency_hierarchy}). Expectedly, the error rates increase to approximately half of all bits being incorrect when $T_Z \approx T_{tr}$ for each sensor and device, as $T_Z$ will always override $T_C$. Finally, it is observed that virtually all bits are received incorrectly on average when $T_Z \approx T_{end}$. The third application's sampling period is equal to the termination frequency, immediately closing the channel; $Recv$ and $Trn$ cannot transmit any bits in this scenario.

\section{Using Multiplexing for Coarse-grained Application Profiling}
\label{sec:application_fingerprinting}

During the course of this work, a second attack vector was discovered that would enable a malicious application to learn information about a sensor-enabled victim app.  We identified that applications were likely to access certain sensors at particular sampling rates. For example, accelerometers are used ubiquitously as a stabilisation mechanism in games for aiming weapons, directing first-person viewpoints, and detecting screen orientation changes~\cite{android:motion_sensors}. Based on this, we developed a two-part study to evaluate the extent to which a malicious application can detect sampling periods used by sensor-enabled victim apps.  This consisted of, firstly, detecting Android Sensor SDK sampling period constants under different sensors and devices.  The Android Sensors SDK provides developers with pre-defined sampling periods in order to minimise battery consumption for several use cases~\cite{android:sensor_manager}. These correspond to: gaming controllers (\texttt{SENSOR\_DELAY\_GAME}, 20ms); UI interactions, e.g.\ gesture recognition (\texttt{SENSOR\_DELAY\_UI}, 60ms); and detecting screen orientation (\texttt{SENSOR\_DELAY\_NORMAL}, 200ms). The fourth, \texttt{SENSOR\_DELAY\_FASTEST}, polls at the maximum supported frequency. In the second part of the study, we evaluated the extent to which sensor-based interactions can be detected using the top 250 Android applications~\cite{android:rank}.

\subsection{Attack Design}

The threat model assumes a malicious app, $M$, that wishes to detect the use of sensor-enabled games, UI interactions, or to directly identify a victim app, $V$. The goal is for $M$ to perform this without requiring additional permissions that may reveal its intentions upon installation. The high-level attack approach follows three steps:
\begin{enumerate}
    \item \emph{Malicious Set-up}: $M$ registers listeners for one or more sensors using their lowest supported sampling frequency simultaneously.
    \item \emph{Victim Execution}: Unwittingly, $V$ registers another listener at another frequency for the same sensor, whether that is a pre-defined period, e.g.\ \texttt{SENSOR\_DELAY\_GAME}, or a custom frequency. Due to multiplexing, one of $M$'s signals is modulated to the higher frequency registered by $V$. $V$ may use the received measurements for its intended purpose before de-registering the listener. 
    \item \emph{Malicious Execution}: Concurrently, $M$ detects $V$'s sampling period and the sensor used, and logs this information to file.
\end{enumerate} 

\begin{figure}
    \centering
    \includegraphics[width=\linewidth]{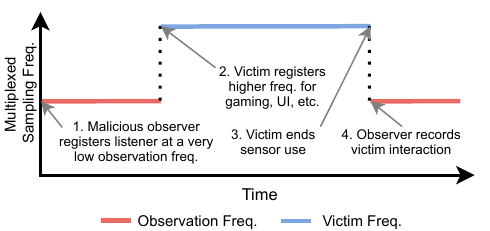}
    \caption{Application profiling overview.}
    \label{fig:fingerprinting}
\end{figure}

This process is illustrated in Figure~\ref{fig:fingerprinting}, with an optional fourth state for recording the interaction. (Note: \ballnumber{1} and \ballnumber{2} are interchangeable. If \ballnumber{2} is performed first, then the faster frequency used by $V$ will already be returned to $M$ when it requests values at an extremely slow frequency.)

\subsection{Implementation}
\label{sec:finger_imp}

The two apps from \S\ref{sec:covert_channels} were modified for inferring the victim sampling frequencies from a malicious observation app. Unlike \S\ref{sec:covert_channels}, however, it cannot be assumed that a particular sensor will be used by a victim app \emph{a priori}. This raises the question: how can $M$ detect multiplexed sampling period changes of an arbitrary sensor used by $V$? To address this, we instrumented $M$ to register sensor listeners for \emph{all} six continuous sensors concurrently, which is valid behaviour under the Android framework. Note that the inferred periods of each sensor must be maintained independently. 

In our test-bed, we implemented $V$ to register a random sensor at one of the aforementioned pre-defined sampling constants from the Android Sensor SDK. This was triggered by user input. The time that $V$ activated the sensor and the sampling period were stored for later analysis. Upon detection, $M$ stored the detected sensor, inferred sampling period, and the associated timestamp.

\subsection{Evaluation}
\label{sec:fingerprint_eval}

This section evaluates the accuracy and latency of the proposed profiling method.

\subsubsection{Methodology}

Initial trials were conducted where $M$ was set to run persistently in the background. This created signals for all six target sensors in parallel using their maximum supported frequencies. Next, $V$ was opened for 25 trials per sensor, device, and sampling period.  Like our covert channel, many of the actual inferred frequencies did not strictly reflect those specified in the Android SDK~\cite{android:sensors_overview}.  For two devices (Poco F1 and Pixel 4A), the observed periods still broadly reflected those specified Android SDK, albeit within a $\pm$10\% error band. Despite this, the observed rates were still individually distinguishable from a long-period carrier signal in most cases. 

It also became apparent that $M$ could not always discriminate between the signals transmitted by $V$ on certain devices. That is, $M$ could not distinguish between \texttt{SensorManager} sampling period constants used by $V$ when multiplexed into $M$'s carrier signal. For instance, the \texttt{SENSOR\_DELAY\_NORMAL} period on the Xiaomi F1's AC and GY sensors \underline{\textit{is}} the maximum supported period. Consequently, these periods used by $V$ cannot be distinguished from $M$'s carrier frequency. Moreover, the Moto G5's single-frequency reporting caused issues once more: the GR, LA, and RV sensors could not be distinguished from the maximum period. As a consequence, the sensors for this device were disregarded from further study.  In total, 20/24 (83.3\%) cases could be successfully detected on the Xiaomi Poco F1 and Google Pixel 4A. Only 6/20 (30\%) of cases were detectable with the Motorola Moto G5. A breakdown of distinguishable cases is shown in Tables~\ref{tab:pocof1}--\ref{tab:motog5}.

\begin{table}[t!]
\centering
\caption{Actual observed sampling periods (ms) for each \texttt{SensorManager} constant (\emph{Xiaomi Poco F1}).}
\begin{threeparttable}
\begin{tabular}{r|ccccc}
\toprule
\textbf{Sensor} & \textbf{Fastest} & \textbf{Game} & \textbf{UI} & \textbf{Normal}& \textbf{Max.} \\ \midrule
\textbf{AC} & \cellcolor{green!40}2.484 & \cellcolor{green!40}19.83 & \cellcolor{green!40}66.67   &\cellcolor{red!40}198.6 & 198.6 \\
\textbf{GR} &  \cellcolor{green!40}4.961 & \cellcolor{green!40}19.89 & \cellcolor{green!40}79.41 & \cellcolor{red!40}198.7 & 198.6  \\
\textbf{GY} & \cellcolor{green!40}4.968 & \cellcolor{green!40}19.87 & \cellcolor{green!40}66.67 & \cellcolor{red!40}198.6 & 198.6 \\
\textbf{LA} & \cellcolor{green!40}4.961 & \cellcolor{green!40}19.89 & \cellcolor{green!40}79.48 & \cellcolor{red!40}198.6 & 198.6 \\
\textbf{MF} & \cellcolor{green!40}10.0 &  \cellcolor{green!40}20.00 & \cellcolor{green!40}66.67 & \cellcolor{green!40}200.0 & 1000   \\
\textbf{RV} & \cellcolor{green!40}4.166 & \cellcolor{green!40}16.64 & \cellcolor{green!40}55.59 & \cellcolor{green!40}166.6 & 833.3\\ \bottomrule
\end{tabular}
\begin{tablenotes}
\item \textit{\textbf{Legend}} --- Green: distinguishable using the max. period as the reference period; Red: indistinguishable. 
\end{tablenotes}
\end{threeparttable}
\label{tab:pocof1}
\end{table}
\begin{table}[t]
\centering
\caption{Actual observed sampling periods (ms) for each \texttt{SensorManager} constant (\emph{Google Pixel 4A}).}
\begin{threeparttable}
\begin{tabular}{r|ccccc}
\toprule
\textbf{Sensor} & \textbf{Fastest} & \textbf{Game} & \textbf{UI} & \textbf{Normal}& \textbf{Max.} \\ \midrule
\textbf{AC} & \cellcolor{green!40}2.346 & \cellcolor{green!40}18.77 & \cellcolor{green!40}65.70   &\cellcolor{green!40}197.1 & 976.2 \\
\textbf{GR} &  \cellcolor{green!40}4.693 & \cellcolor{green!40}18.77 & \cellcolor{green!40}65.70 & \cellcolor{red!40}197.1 & 197.1  \\
\textbf{GY} & \cellcolor{green!40}2.346 & \cellcolor{green!40}18.77 & \cellcolor{green!40}65.70 & \cellcolor{green!40}197.1 & 976.2 \\
\textbf{LA} & \cellcolor{red!40}14.08$\star$ & \cellcolor{green!40}18.77 & \cellcolor{green!40}65.70 & \cellcolor{red!40}197.1 & 197.1 \\
\textbf{MF} & \cellcolor{green!40}9.956 &  \cellcolor{green!40}19.58 & \cellcolor{green!40}66.67 & \cellcolor{green!40}195.9 & 979.5   \\
\textbf{RV} & \cellcolor{green!40}4.693 & \cellcolor{green!40}18.77 & \cellcolor{green!40}65.70 & \cellcolor{red!40}197.1 & 197.1\\ \bottomrule
\end{tabular}
\begin{tablenotes}
\item $\star$: Erroneously returned the \texttt{SENSOR\_DELAY\_GAME} period when \texttt{SENSOR\_DELAY\_FASTEST} was used.
\end{tablenotes}
\end{threeparttable}
\label{tab:pixel4a}
\end{table}
\begin{table}[t!]
\centering
\caption{Actual observed sampling periods (ms) for each \texttt{SensorManager} constant (\emph{Motorola Moto G5}).}
\begin{threeparttable}
\begin{tabular}{r|ccccc}
\toprule
\textbf{Sensor} & \textbf{Fastest} & \textbf{Game} & \textbf{UI} & \textbf{Normal}& \textbf{Max.} \\ \midrule
\textbf{AC} & \cellcolor{green!40}9.934 & \cellcolor{green!40}19.98 & \cellcolor{green!40}59.68   &\cellcolor{red!40}198.6 & 198.6 \\
\textbf{GR} &  \cellcolor{red!40}5.026$\star$ & \cellcolor{red!40}9.893 & \cellcolor{red!40}9.941 & \cellcolor{red!40}9.890 & 9.915  \\
\textbf{GY} & \cellcolor{green!40}4.944 & \cellcolor{green!40}19.87 & \cellcolor{green!40}59.59 & \cellcolor{red!40}198.6 & 198.6 \\
\textbf{LA} & \cellcolor{red!40}4.958$\star$ & \cellcolor{red!40}9.893 & \cellcolor{red!40}9.863 & \cellcolor{red!40}9.879 & 9.901 \\
\textbf{RV} & \cellcolor{yellow!40}4.950$\Diamond$ & \cellcolor{red!40}9.906 & \cellcolor{red!40}10.03 & \cellcolor{red!40}9.926 & 9.918\\ \bottomrule
\end{tabular}
\begin{tablenotes}
\item $\star$: Failed to return \texttt{SENSOR\_DELAY\_FASTEST} when multiplexed into the maximum period.
\item $\Diamond$: GY was modulated when RV was activated. (Note: RV is a virtual sensor that uses GY as input.)
\end{tablenotes}
\end{threeparttable}
\label{tab:motog5}
\end{table}

\subsubsection{Identifying Android Sampling Constants}
\label{sec:fingerprinting_results}

\begin{figure}[t!]
    \centering
    \includegraphics[width=\linewidth]{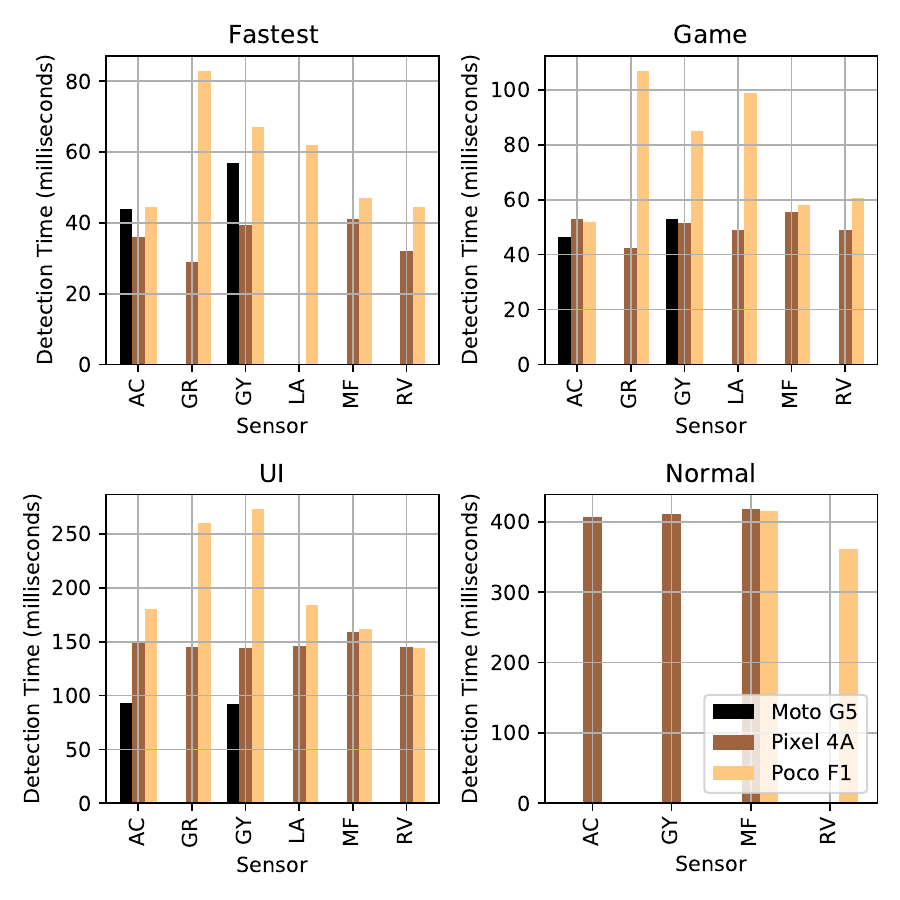}
    \caption{Average profiling detection latency for each \texttt{SensorManager} constant, sensor, and device.}
    \label{fig:fingerprint_detection}
\end{figure}

Following the initial results, we conducted further experiments of 100 additional trials per sensor, per sampling period, and per device. This produced 4,600 interactions in total between $V$ and $M$. In total, $M$ was able to correctly detect $V$'s chosen sensor and sampling period in all observed cases. The average detection latencies for each sensor, device, and constant are given in Figure~\ref{fig:fingerprint_detection}.  On average, a given sensor and sampling constant was detected by $M$ in $<$500ms.   Notably, the detection latency was reduced as $V$'s chosen sampling period was lower (faster). This is a natural result of our sampling period inferencing method, which uses sequential sensor events. When $V$ requests measurements at a higher frequency, then measurements are multiplexed and inferred by $M$ at a faster rate.

In the best cases, the detection latency was reduced to $<$80ms when the victim used the sensor's fastest sampling rate. In a small number of cases, detection latency reached under half (40ms), e.g.\ the Pixel 4A's AC and RV sensors. At the opposite end, $V$'s sampling period was inferred in $\sim$400ms where \texttt{SENSOR\_DELAY\_NORMAL} was used. The \texttt{SENSOR\_DELAY\_UI} constant could be detected in $<$250ms for all devices and sensors, and $<$150ms for the Pixel 4A and Moto G5. Lastly, \texttt{SENSOR\_DELAY\_GAME} was detected in $<$100ms, and $<$60ms for the Pixel 4A and Moto G5 on average. In general, the detection latency is device- and sensor-dependent. The detection latency was largely consistent for the Pixel 4A. In contrast to the covert channel, the Moto G5 performed relatively well for the remaining sensors. The Poco F1 performed with reduced consistency; in particular, the GR and GY sensors required more time to detect $V$'s sampling period.  Yet, in all cases, this could be achieved in under half a second on our test devices.

\subsubsection{Evaluating Real-World Applications}
\label{sec:evaluating-real-world-apps}
The previous section showed how particular sampling constants used by a victim application can be identified by a malicious observer. To extend these results, we investigated the top 250 most popular Android applications (as of February 2022, according to AndroidRank~\cite{android:rank}). In this study, the malicious application, $M$, was launched that registered listeners for all supported device sensors at the slowest possible frequency. Next, each application from the top 250 was launched. We proceeded beyond main menus, login dialogs, and other intermediate interfaces to reach the main activity where sensors were utilised (if any). If no sensors were used, then the application was discarded from further consideration. In cases where sensors were used, the interaction was logged to file by $M$, which was subsequently retrieved for off-line analysis. 

In total, our malicious observer detected 57/250 apps (22.8\%) where sensors were used. The majority of these were games (38; 66.7\%), followed by food and drink (3; 5.2\%), maps and navigation (3), photography (3), music and audio (2; 3.5\%), travel and local (2), art and design (1; 1.7\%), finance (1), health and fitness (1), productivity (1), shopping (1), and social (1). This is illustrated in Figure~\ref{fig:app_breakdown}. Comprehensive results are given in Appendix~\ref{appendix:fingerprinted_apps} in Table~\ref{tab:app_fingerprinting_mobile}. Of the gaming applications, the most common sub-categories were action (18),  racing (7), adventure (2), arcade (2), casual (2), trivia (2), sports (2), board (1), puzzle (1), and simulation (1).

\begin{figure}
    \centering
    \includegraphics[width=\linewidth]{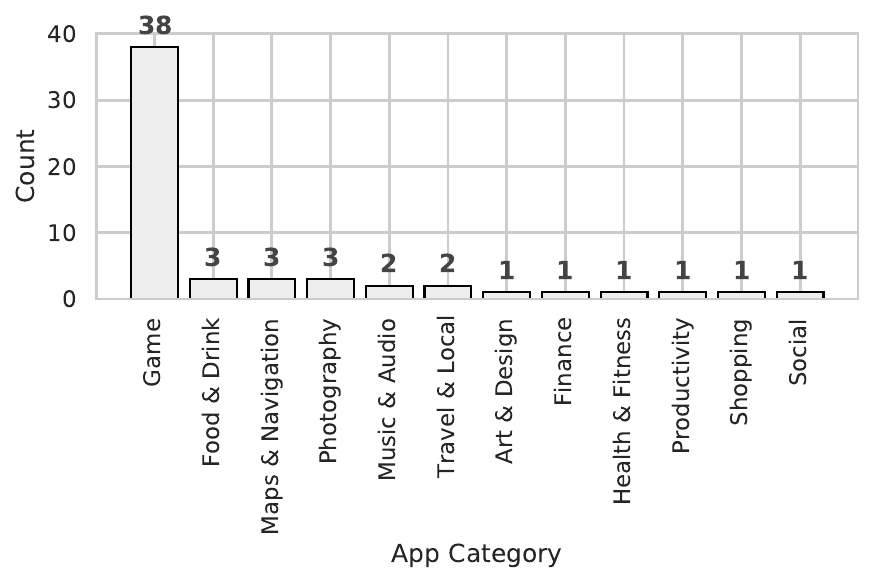}
    \caption{Category breakdown of detected applications.}
    \label{fig:app_breakdown}
\end{figure}

Following our analysis, we found that the vast majority of apps used the AC sensor (53/57; 92.9\%), primarily at the \texttt{SENSOR\_DELAY\_GAME} frequency (33/53; 62.3\%).  This is not surprising: accelerometers are supported by \emph{``most''} Android devices for implementing tilting, shaking, rotation and swinging-based interactions, particularly in games~\cite{android:motion_sensors}. What was surprising was the large number of sensors and non-standard sampling rates used by sophisticated apps, e.g.\ Pok\'emon Go---an augmented reality (AR) game by Niantic, Inc. This particular application used six different sensors---AC, GR, MF, LA, and RV---with a non-standard 10ms sampling period for the AC sensor. Indeed, this application used a large and unique combination of sensors and sampling rates. Interestingly, we found unique sensor-sampling combinations were used by 13/57 (22.8\%) of detected apps. This information may enable a malicious actor to precisely profile individual applications. Comparatively, the remaining applications used conflicting parameters (44/57; 77.2\%). That is, the sensor(s) and sampling period(s) were used by at least one other application.  While individual identification is challenging here, an attacker may still be able to deduce the victim's high-level nature. For instance, we observed that all but one game (37/38; 97.4\%) used the AC sensor at the \texttt{SENSOR\_DELAY\_GAME} frequency or greater. This increased to 100\% of apps within some sub-categories (racing and simulation). In addition, we recognised that apps which relied on real-world navigation, e.g.\ location and heading, typically utilised both the AC and MF sensors at a high frequency (Gojek, Google Fit, Google Maps, Grab, Pok\'emon Go, Uber, Waze, and Yandex Go).

\section{Security Evaluation}
\label{sec:analysis}

To address sensor-related security issues (see \S\ref{sec:related_work}), Android 9 introduced updates for visibly indicating when high-frequency, long-polling sensors are active. In this section, these changes are examined and how they are unsatisfactory for addressing our attacks. We discuss UI- and system-level countermeasures, and offer recommendations.

\subsection{Android 9 Changes}

From Android 9 (API level 28), applications cannot use background services to access sensors in continuous reporting mode. If this is attempted, then sensor events are not returned to the requesting application. Instead, apps must use foreground services that are visible to the user using a taskbar notification. This is to raise the user's awareness of on-going background processing. The notification is removed only when the service is stopped. 
Therefore, Android application manifest files must contain the \texttt{FOREGROUND\_SERVICE} permission in order to access sensor services. Unfortunately, this leads to the first issue: \texttt{FOREGROUND\_SERVICE} is a `normal' permission. That is, according to the Android documentation, the associated \emph{``data and actions present very little risk to the user's privacy''}~\cite{android:permissions}. Android OS automatically grants normal permissions to requesting applications without user input. This compares with `dangerous' permissions---e.g.\ accessing SMS messages, call logs, and location data---that requires user confirmation through a dialog prompt~\cite{android:dangerous}. Put otherwise, a malicious application can access sensor services to mount our attacks without requiring explicit user approval.

\begin{figure}
    \centering
    \includegraphics[width=0.67\linewidth]{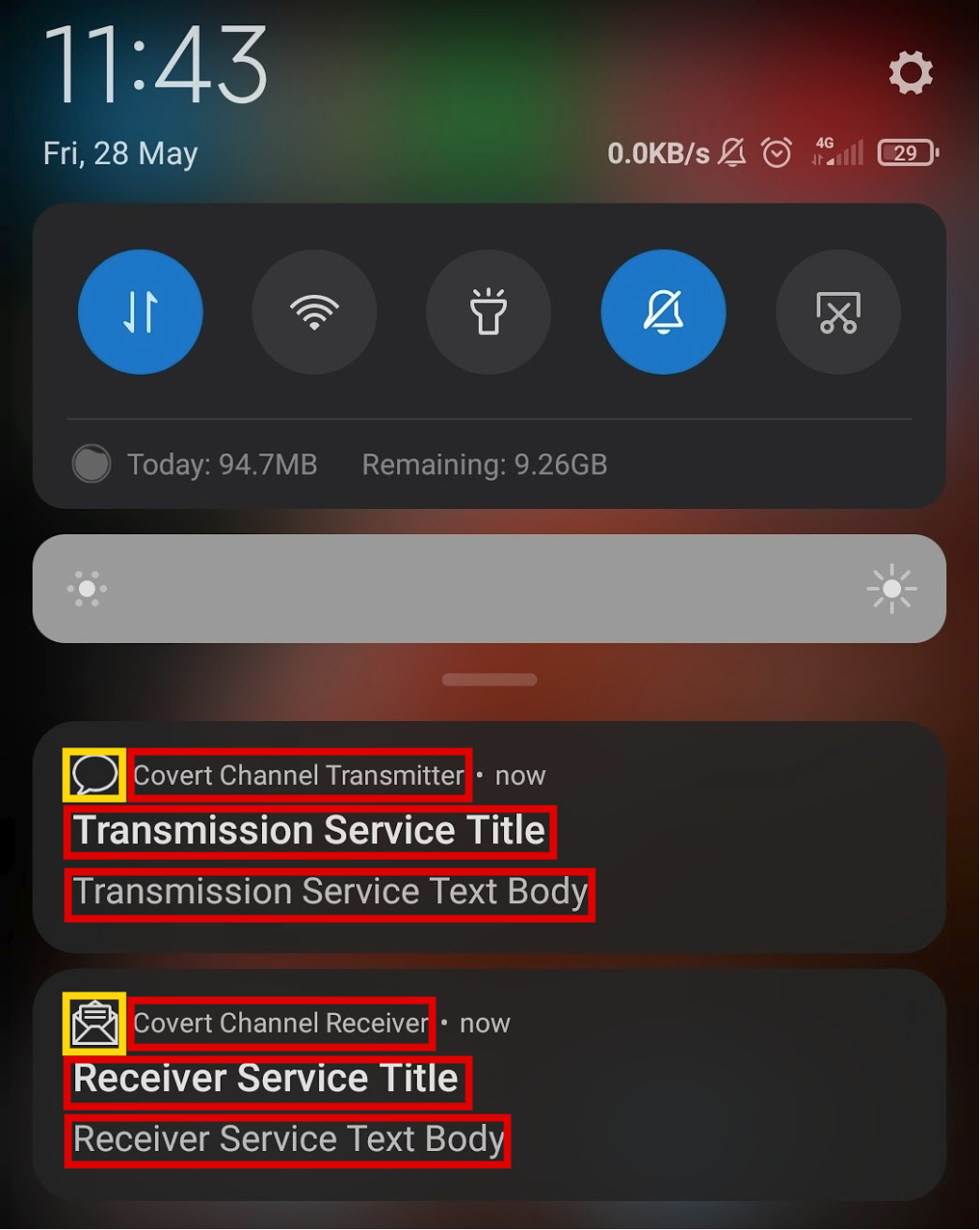}
    \caption{Covert channel foreground notifications. Yellow regions: customisable images; red: customisable text areas.}
    \label{fig:pocof1_annotated}
\end{figure}

Secondly, foreground service notifications are heavily customisable (Figure~\ref{fig:pocof1_annotated}). The notification heading, text body, and icon can all be changed with minimal restrictions by a malicious developer. It is conceivable that an attacker could manipulate these properties in order to masquerade as a seemingly legitimate application.  A potential UI countermeasure is to require applications to display the full set of active registered sensor listeners in their foreground service notification. This would indicate any unusual sensor usage that strays beyond the application's expected remit. Yet, using the technical names of sensors is likely to prompt user confusion. Referring to the \emph{class} of sensors in use, e.g.\ `motion' or `position', may be more appropriate. However, a usability investigation for indicating potentially dangerous sensors to users is warranted.

\subsection{Recommendations}

From the design of the Sensors SDK, a robust countermeasure is to remove the ability for application developers to control the sampling rate through the \texttt{registerListener} method. Instead, the appropriate sampling rate could be inferred and returned to the application at an operating system level.

\vspace*{5pt}
\fbox{\begin{minipage}{225pt}
\textbf{Recommendation 1: Consider removing the ability to directly set sensor sampling periods.}\\
By adopting this measure, malicious apps are unable to establish carrier signals at a desired frequency, thereby preventing both presented attacks.
\end{minipage}}
\vspace*{5pt}

Both attacks leverage the design choice of returning measurements at the fastest sampling rate when multiple apps request different rates. On Apple iOS devices, sensor measurements are acquired  through the SensorKit for general sensors~\cite{appleios:sensorkit} (Apple iOS 14.0+) and the Core Motion framework~\cite{appleios:coremotion} (iOS 4.0+) for motion sensors, e.g.\ AC and GY. We developed two applications that accessed sensors through these APIs at different sampling frequencies. On an Apple iPhone 13, 13 Pro, and the official emulator, sensor measurements were returned independently to both applications at their requested frequencies. Thus, we were unable to replicate our attacks that utilise frequency-shift keying. (Despite best efforts, we cannot precisely elucidate \emph{how} multiple frequency requests are handled due to the proprietary, closed-source nature of the Apple sensor framework). Similar changes could be implemented on Android as a countermeasure, i.e.\ enforcing that requested sampling periods are the ones received by applications. However, this requires fundamental changes to the measurement delivery mechanism in the Android sensor stack. 

\vspace*{5pt}
\fbox{\begin{minipage}{225pt}
\textbf{Recommendation 2: Enforce requested sampling periods on a per-app basis.}\\
This prevents both attacks by stopping malicious apps from exploiting software multiplexing using higher frequency signals registered by other apps. 
\end{minipage}}

\subsection{Limitations}

The attacks perform optimally where malicious apps use unused or under-used sensors. The covert channel (\S\ref{sec:covert_channels}) cannot be mounted when a third app is already using a sensor at the highest frequency, preventing the receiver and transmitter from using the necessary frequency bands. In these cases, the attacker is best suited to utilising rarely used sensors, such as gravity and linear acceleration sensors, where available. Note that a malicious application can detect when sensor(s) are already being used at its fastest rate by a third application. This can be achieved by registering multiple sensors and comparing the inferred rate with the highest supported frequency from the \texttt{SensorManager} class. It is also worth noting that our channel does not have the same throughput as hardware-oriented channels. It is common for cache-based cover channels to have extremely high throughput, such as 500 kbps~\cite{liu2015last} and 751 bps~\cite{maurice2015c5} using L3 cache activity. Covert channels based on RAM have similarly high widths: 333 bps for a memory bus lock channel~\cite{semal2020one}, 69--729 bps for a RAM controller channel \cite{semal2020leaky} and 411 kbps--2 Mbps for the DRAMA channel by Pessl et al.~\cite{pessl2016drama}. In general, however, micro-architectural and RAM-based channels tend to require privileged execution to access system instructions and possess architectural dependencies (e.g.\ only X86~\cite{semal2020leaky,semal2020one,maurice2015c5}). Comparatively, while our software-based channel is slower (5.10--9.62 bps), we stress that our work imposes very few assumptions: it is CPU- and SoC-agnostic, and uses standard interfaces available to any Android app developer. 

As was discussed in \S\ref{sec:evaluating-real-world-apps}, the profiling approach is best suited to apps with rarely or uniquely used sensor/frequency combinations. Reduced utility is observed if the target app uses a common combination. For example, the AC sensor with the \texttt{SENSOR\_DELAY\_GAME} frequency is a common configuration for apps in the `Game' category. (Note that this still confers some information to an attacker, i.e.\ the user is likely playing a game). In this scenario, the potential subset of applications being executed is reduced. This subset becomes particularly tight when the target application uses a unique sensor/sampling combination.

\section{Conclusion}
\label{sec:conclusion}

In this paper, we presented two software-controlled side-channel methods that arise from insecure design choices in mobile sensor stacks. Specifically, we show how software-based multiplexing can result in the creation of reliable FSK-based covert channels and the ability to profile sensor-enabled applications. Neither approach imposes special requirements---for example, a rooted handset or kernel-mode access---beyond the standard Sensor SDK made available to application developers. 

In \S\ref{sec:covert_channels}, a spectral covert channel was developed for transmitting arbitrary bit-strings between same-device applications in a way that bypassed IPC security mechanisms. This was achieved using a carrier frequency established by a receiver app, which was modulated by a transmitted app that requested a higher sensor sampling frequency. It was shown how many physical and virtual continuous sensors could be used for this purpose, including accelerometers, gyroscopes, and magnetic field sensors, with low error rates. Indeed, the transmission of low-resolution yet legible single-bit images was possible at up to $\sim$10 bps.

In \S\ref{sec:application_fingerprinting}, a variant of this technique was developed for profiling sensor-enabled applications at a coarse-grained level. This involved a malicious app that generated multiple, simultaneous sensor signals using their lowest supported frequency. Victim apps which used the same sensor(s) at a higher frequency thus triggered modulations in this signal. It was shown how this could be used for detecting particular interactions (e.g.\ UI interactions, screen orientation changes, and gaming activity) with low latency (30--400ms). Moreover, an analysis of the top 250 Android applications showed the activity of 57 (22.8\%) were detected using our approach. It was shown how some highly downloaded applications exhibited unique profiles. This was due to using unique sensor and sampling frequency combinations, which can be passively detected by a malicious application.

For both approaches, all evaluation devices from different OEMs were vulnerable, potentially affecting millions of devices worldwide.  In \S\ref{sec:analysis}, we suggested two recommendations: \ballnumber{1} removing the ability for developers to set precise sensor sampling periods; and \ballnumber{2} enforcing the requested sampling periods by removing the multiplexing functionality. Unfortunately, both solutions require design changes to the Android sensor stack.


\ifCLASSOPTIONcompsoc
  \section*{Acknowledgments}
\else
  \section*{Acknowledgment}
\fi

This work received funding from the European Union's Horizon 2020 research and innovation programme under grant agreement No.\ 883156 (EXFILES). 
\bibliographystyle{plain}
\bibliography{bib}

\begin{thebibliography}{10}

\bibitem{android:batching}
{Android Open Source Project}.
\newblock Batching, 2021.
\newblock \url{https://source.android.com/devices/sensors/batching}.

\bibitem{android:fmqs}
{Android Open Source Project}.
\newblock Fast message queues {(FMQs)}, 2021.
\newblock \url{https://source.android.com/devices/architecture/hidl/fmq}.

\bibitem{android:dangerous}
{Android Open Source Project}.
\newblock Manifest.permission, 2021.
\newblock \url{https://developer.android.com/reference/android/Manifest.permission}.

\bibitem{android:motion_sensors}
{Android Open Source Project}.
\newblock Motion sensors, 2021.
\newblock \url{https://developer.android.com/guide/topics/sensors/sensors_motion}.

\bibitem{android:permissions}
{Android Open Source Project}.
\newblock Permissions on {Android}, 2021.
\newblock \url{https://developer.android.com/guide/topics/permissions/overview\#normal-dangerous}.

\bibitem{android:hal2}
{Android Open Source Project}.
\newblock Sensor {HAL2}, 2021.
\newblock \url{https://source.android.com/devices/sensors/sensors-hal2}.

\bibitem{android:sensors_overview}
{Android Open Source Project}.
\newblock Sensor overview, 2021.
\newblock \url{https://developer.android.com/guide/topics/sensors/sensors_overview}.

\bibitem{android:sensor_stack}
{Android Open Source Project}.
\newblock Sensor stack, 2021.
\newblock \url{https://source.android.com/devices/sensors/sensor-stack}.

\bibitem{android:sensor_manager}
{Android Open Source Project}.
\newblock {SensorManager}, 2021.
\newblock \url{https://developer.android.com/reference/android/hardware/SensorManager}.

\bibitem{android:rank}
{AndroidRank}.
\newblock Open {Android} market data, 2022.
\newblock \url{https://www.androidrank.org/}.

\bibitem{appleios:coremotion}
{Apple, Inc.}
\newblock {Core Motion -- Apple Developer Documentation}, 2023.
\newblock \url{https://developer.apple.com/documentation/coremotion}.

\bibitem{appleios:sensorkit}
{Apple, Inc.}
\newblock {SensorKit -- Apple Developer Documentation}, 2023.
\newblock \url{https://developer.apple.com/documentation/sensorkit}.

\bibitem{bartolini2016capacity}
Davide~B Bartolini, Philipp Miedl, and Lothar Thiele.
\newblock On the capacity of thermal covert channels in multicores.
\newblock In {\em Proceedings of the 11th European Conference on Computer Systems}, pages 1--16, 2016.

\bibitem{block2017autonomic}
Kenneth Block, Sashank Narain, and Guevara Noubir.
\newblock An autonomic and permissionless android covert channel.
\newblock In {\em 10th ACM Conf.\ on Security and Privacy in Wireless and Mobile Networks}, 2017.

\bibitem{cabuk2004ip}
Serdar Cabuk, Carla~E Brodley, and Clay Shields.
\newblock {IP} covert timing channels: {Design} and detection.
\newblock In {\em 11th ACM Conference on Computer and Communications Security}, pages 178--187, 2004.

\bibitem{cai2011touchlogger}
Liang Cai and Hao Chen.
\newblock Touchlogger: Inferring keystrokes on touch screen from smartphone motion.
\newblock {\em HotSec}, 11(2011):9, 2011.

\bibitem{cho2018prime}
Haehyun Cho, Penghui Zhang, Donguk Kim, Jinbum Park, Choong-Hoon Lee, Ziming Zhao, Adam Doup{\'e}, and Gail-Joon Ahn.
\newblock {Prime+Count}: Novel cross-world covert channels on {ARM TrustZone}.
\newblock In {\em 34th Annual Computer Security Applications Conference}, 2018.

\bibitem{goel2012gripsense}
Mayank Goel, Jacob Wobbrock, and Shwetak Patel.
\newblock Gripsense: using built-in sensors to detect hand posture and pressure on commodity mobile phones.
\newblock In {\em 25th ACM Symposium on User Interface Software and Technology}, pages 545--554, 2012.

\bibitem{gruss2016flush}
Daniel Gruss, Cl{\'e}mentine Maurice, Klaus Wagner, and Stefan Mangard.
\newblock {Flush+Flush}: a fast and stealthy cache attack.
\newblock In {\em International Conference on Detection of Intrusions and Malware, and Vulnerability Assessment}, pages 279--299. Springer, 2016.

\bibitem{gupta2016continuous}
Hari~Prabhat Gupta, Haresh~S Chudgar, Siddhartha Mukherjee, Tanima Dutta, and Kulwant Sharma.
\newblock A continuous hand gestures recognition technique for human-machine interaction using accelerometer and gyroscope sensors.
\newblock {\em IEEE Sensors Journal}, 16(16):6425--6432, 2016.

\bibitem{gurulian2017effectiveness}
Iakovos Gurulian, Carlton Shepherd, Eibe Frank, Konstantinos Markantonakis, Raja~Naeem Akram, and Keith Mayes.
\newblock On the effectiveness of ambient sensing for detecting {NFC} relay attacks.
\newblock In {\em 16th IEEE International Conference on Trust, Security and Privacy in Computing and Communications}, pages 41--49. IEEE, 2017.

\bibitem{halevi2012secure}
Tzipora Halevi, Di~Ma, Nitesh Saxena, and Tuo Xiang.
\newblock Secure proximity detection for {NFC} devices based on ambient sensor data.
\newblock In {\em European Symposium on Research in Computer Security}, pages 379--396. Springer, 2012.

\bibitem{jiang2016complete}
Zhen~Hang Jiang, Yunsi Fei, and David Kaeli.
\newblock A complete key recovery timing attack on a {GPU}.
\newblock In {\em IEEE International Symposium on High Performance Computer Architecture}. IEEE, 2016.

\bibitem{krosche2018dpid}
Robert Kr{\"o}sche, Kashyap Thimmaraju, Liron Schiff, and Stefan Schmid.
\newblock I {DPID} it my way! {A} covert timing channel in software-defined networks.
\newblock In {\em IFIP Networking Conference and Workshops}, pages 217--225. IEEE, 2018.

\bibitem{lin2015designing}
Yuqi Lin, Saif~UR Malik, Kashif Bilal, Qiusong Yang, Yongji Wang, and Samee~U Khan.
\newblock Designing and modeling of covert channels in operating systems.
\newblock {\em IEEE Transactions on Computers}, 2015.

\bibitem{lipp2016armageddon}
Moritz Lipp, Daniel Gruss, Raphael Spreitzer, Cl{\'e}mentine Maurice, and Stefan Mangard.
\newblock Armageddon: Cache attacks on mobile devices.
\newblock In {\em 25th {USENIX} Security Symposium ({USENIX} Security)}, pages 549--564, 2016.

\bibitem{liu2015last}
Fangfei Liu, Yuval Yarom, Qian Ge, Gernot Heiser, and Ruby~B Lee.
\newblock Last-level cache side-channel attacks are practical.
\newblock In {\em IEEE Symposium on Security and Privacy}, pages 605--622. IEEE, 2015.

\bibitem{long2018improving}
Zijun Long, Xiaohang Wang, Yingtao Jiang, Guofeng Cui, Li~Zhang, and Terrence Mak.
\newblock Improving the efficiency of thermal covert channels in multi-/many-core systems.
\newblock In {\em Design, Automation and Test in Europe}, pages 1459--1464. IEEE, 2018.

\bibitem{masti2015thermal}
Ramya~Jayaram Masti, Devendra Rai, Aanjhan Ranganathan, Christian M{\"u}ller, Lothar Thiele, and Srdjan Capkun.
\newblock Thermal covert channels on multi-core platforms.
\newblock In {\em 24th {USENIX} Security Symposium ({USENIX} Security)}, pages 865--880, 2015.

\bibitem{matyunin2016covert}
Nikolay Matyunin, Jakub Szefer, Sebastian Biedermann, and Stefan Katzenbeisser.
\newblock Covert channels using mobile device's magnetic field sensors.
\newblock In {\em 21st Asia and South Pacific Design Automation Conference}, pages 525--532. IEEE, 2016.

\bibitem{matyunin2019magneticspy}
Nikolay Matyunin, Yujue Wang, Tolga Arul, Kristian Kullmann, Jakub Szefer, and Stefan Katzenbeisser.
\newblock {MagneticSpy}: Exploiting magnetometer in mobile devices for website and application fingerprinting.
\newblock In {\em Proceedings of the 18th ACM Workshop on Privacy in the Electronic Society}, pages 135--149, 2019.

\bibitem{maurice2015c5}
Cl{\'e}mentine Maurice, Christoph Neumann, Olivier Heen, and Aur{\'e}lien Francillon.
\newblock C5: Cross-cores cache covert channel.
\newblock In {\em International Conference on Detection of Intrusions and Malware, and Vulnerability Assessment}, pages 46--64. Springer, 2015.

\bibitem{michalevsky2014gyrophone}
Yan Michalevsky, Dan Boneh, and Gabi Nakibly.
\newblock Gyrophone: Recognizing speech from gyroscope signals.
\newblock In {\em 23rd {USENIX} Security Symposium ({USENIX} Security)}, pages 1053--1067, 2014.

\bibitem{naghibijouybari2018rendered}
Hoda Naghibijouybari, Ajaya Neupane, Zhiyun Qian, and Nael Abu-Ghazaleh.
\newblock Rendered insecure: {GPU} side channel attacks are practical.
\newblock In {\em ACM Conf.\ on Computer and Communications Security}, 2018.

\bibitem{novak2015physical}
Ed~Novak, Yutao Tang, Zijiang Hao, Qun Li, and Yifan Zhang.
\newblock Physical media covert channels on smart mobile devices.
\newblock In {\em Proceedings of the ACM International Joint Conference on Pervasive and Ubiquitous Computing}, pages 367--378, 2015.

\bibitem{osvik2006cache}
Dag~Arne Osvik, Adi Shamir, and Eran Tromer.
\newblock Cache attacks and countermeasures: the case of {AES}.
\newblock In {\em Cryptographers' Track at the RSA Conference}, pages 1--20. Springer, 2006.

\bibitem{owusu2012accessory}
Emmanuel Owusu, Jun Han, Sauvik Das, Adrian Perrig, and Joy Zhang.
\newblock Accessory: Password inference using accelerometers on smartphones.
\newblock In {\em Proceedings of the 12th Workshop on Mobile Computing Systems and Applications}, pages 1--6, 2012.

\bibitem{pessl2016drama}
Peter Pessl, Daniel Gruss, Cl{\'e}mentine Maurice, Michael Schwarz, and Stefan Mangard.
\newblock {DRAMA}: Exploiting {DRAM} addressing for cross-{CPU} attacks.
\newblock In {\em 25th USENIX Security Symposium}, 2016.

\bibitem{semal2020leaky}
Benjamin Semal, Konstantinos Markantonakis, Raja~Naeem Akram, and Jan Kalbantner.
\newblock Leaky controller: Cross-{VM} memory controller covert channel on multi-core systems.
\newblock In {\em IFIP Conference on ICT Systems Security and Privacy Protection}. Springer, 2020.

\bibitem{semal2020one}
Benjamin Semal, Konstantinos Markantonakis, Keith Mayes, and Jan Kalbantner.
\newblock One covert channel to rule them all: A practical approach to data exfiltration in the cloud.
\newblock In {\em 19th Int'l Conf.\ on Trust, Security and Privacy in Computing and Communications}. IEEE, 2020.

\bibitem{ulz2019sensing}
Thomas Ulz, Markus Feldbacher, Thomas~W Pieber, and Christian Steger.
\newblock Sensing danger: Exploiting sensors to build covert channels.
\newblock In {\em International Conference on Information Systems Security and Privacy}, ICISSP, pages 100--113, 2019.

\bibitem{wang2012gesture}
Xian Wang, Paula Tarr{\'\i}o, Eduardo Metola, Ana~M Bernardos, and Jos{\'e}~R Casar.
\newblock Gesture recognition using mobile phone’s inertial sensors.
\newblock In {\em Distributed Computing and Artificial Intelligence}, pages 173--184. Springer, 2012.

\bibitem{xu2012taplogger}
Zhi Xu, Kun Bai, and Sencun Zhu.
\newblock Taplogger: Inferring user inputs on smartphone touchscreens using on-board motion sensors.
\newblock In {\em Proceedings of the 5th ACM Conference on Security and Privacy in Wireless and Mobile Networks}, pages 113--124, 2012.

\bibitem{yarom2014flush}
Yuval Yarom and Katrina Falkner.
\newblock {Flush+Reload}: A high resolution, low noise, {L3} cache side-channel attack.
\newblock In {\em 23rd {USENIX} Security Symposium ({USENIX} Security)}, pages 719--732, 2014.

\end{thebibliography}

\begin{IEEEbiography}[{\includegraphics[width=1in,height=1.25in,clip,keepaspectratio]{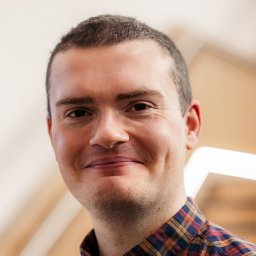}}]{Carlton Shepherd}  received his
Ph.D. in Information Security from the Information Security Group at Royal Holloway, University of London, U.K, and B.S.\ in Computer Science from Newcastle University, U.K. He is a Lecturer in Computing at Newcastle University, U.K., following a Senior Research Fellow position at the Information Security Group at Royal Holloway, University of London. His research interests include trusted execution environments (TEEs) and their applications, and embedded systems security.
\end{IEEEbiography}

\begin{IEEEbiography}[{\includegraphics[width=1in,height=1.25in,clip,keepaspectratio]{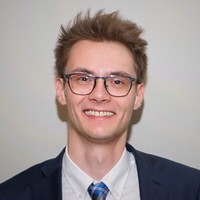}}]{Jan Kalbantner}
received his B.S.\ in Computer Science from Cooperative State University BW, Mosbach, Germany and his M.S.\ in Information Security from Royal Holloway, University of London, U.K., in 2016 and 2019 respectively. Since 2019, he is working towards his Ph.D.\ degree at Royal Holloway, University of London. His research interests include cyber-physical systems, DLT, network security, and secure architectures.
\end{IEEEbiography}

\begin{IEEEbiography}[{\includegraphics[width=1in,height=1.25in,clip,keepaspectratio]{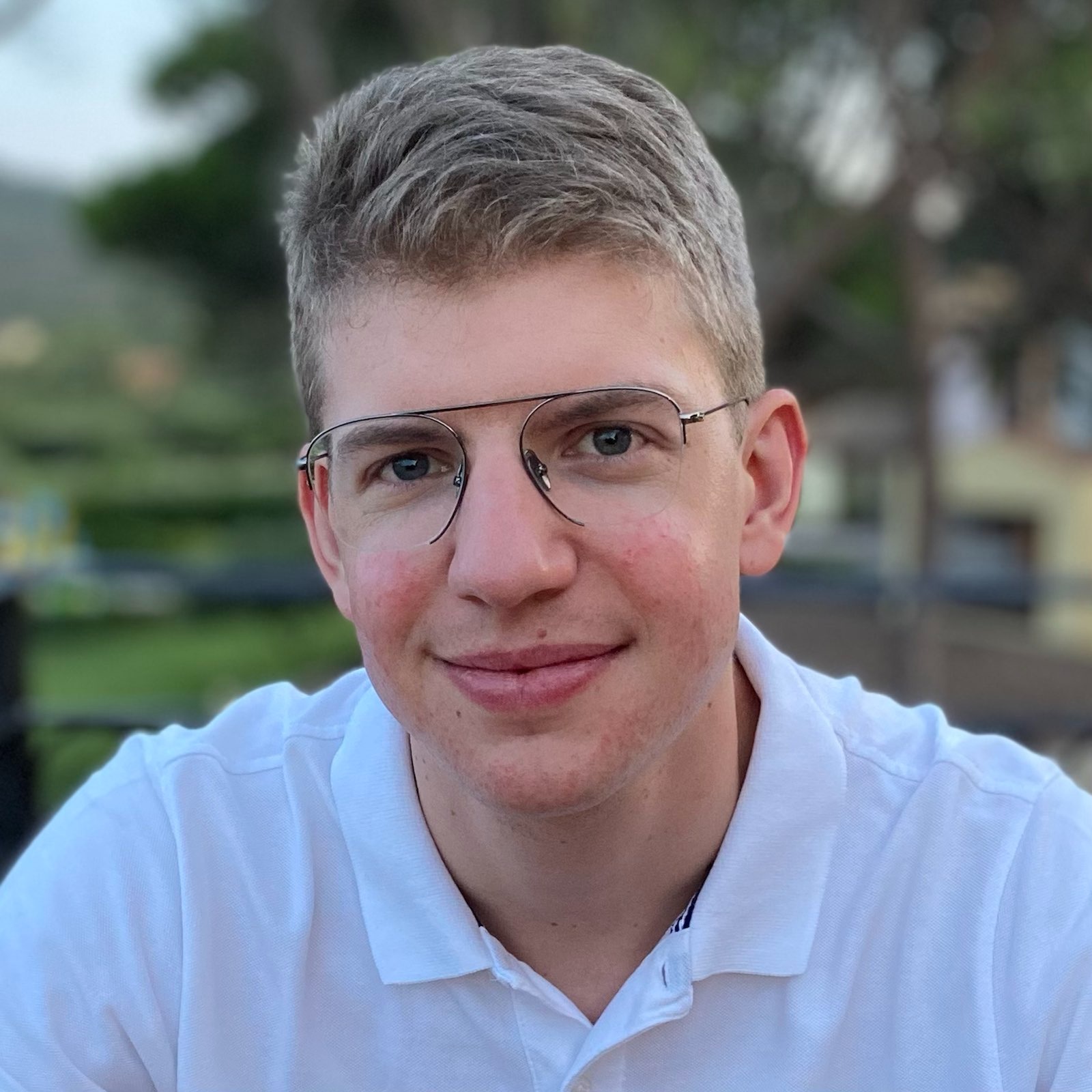}}]{Benjamin Semal}
received his Ph.D.\ in Information Security at Royal Holloway, University of London; M.Eng.\ in Electrical Engineering from Ecole Polytechnique Universitaire of Montpellier, France; and M.S. in robotics from Cranfield University, U.K. He was a hardware security analyst at UL Transaction Security, and now works as a security engineer at SERMA Security \& Safety evaluating point-of-sale devices and cryptographic modules. His research focusses on side-channel attacks for information leakage.
\end{IEEEbiography}

\begin{IEEEbiography}[{\includegraphics[width=1in,height=1.25in,clip,keepaspectratio]{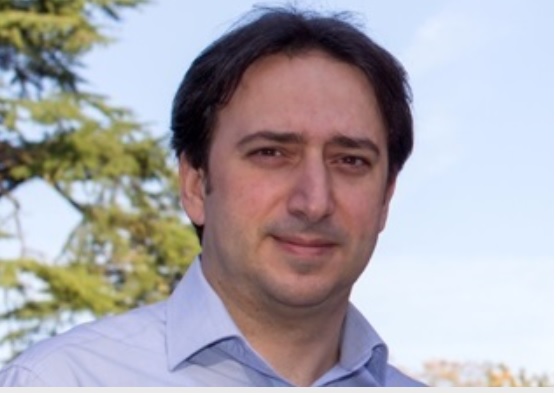}}]{Konstantinos Markantonakis}
received his B.S.\ in Computer Science from Lancaster University, U.K.; and his M.S.\ and Ph.D.\ in Information Security, and M.B.A. in International Management from Royal Holloway, University of London, London, UK. He is currently the Director of the Smart Card and IoT Security Centre. He has co-authored over 190 papers in international conferences and journals. His research interests include smart card security, trusted execution environments, and the Internet of Things.
\end{IEEEbiography}


\appendices

\section{Covert Channel Sampling Parameters}
\label{sec:appendix_params}

This appendix provides the precise parameters used for each device and sensor for establishing FSK-based covert channels. These are shown in Tables~\ref{tab:pixel4a_cc_periods}--\ref{tab:motog5_cc_periods}.

\begin{table}[H]
\centering
\caption{Google Pixel 4A channel sampling periods ($\mu$s).}
\begin{tabular}{@{}r|cccc@{}}
\toprule
\textbf{Sensor} & \textbf{$T_{c}$} & \textbf{$T_{tr}$} & \textbf{$T_{sync}$} & \textbf{$T_{end}$}  \\ \midrule
\textbf{AC} & 10000 & 7500 & 5000 & 2500  \\
\textbf{GR} & 20000 & 15000 & 10000 & 5000  \\
\textbf{GY} & 10000 & 7500 & 5000 & 2500  \\
\textbf{LA} & 50000 & 40000 & 30000 & 20000  \\
\textbf{MF} & 25000 & 20000 & 15000 & 10000  \\
\textbf{RV} & 20000 & 15000 & 10000 & 5000  \\ \bottomrule
\end{tabular}
\label{tab:pixel4a_cc_periods}
\end{table}
\begin{table}[H]
\centering
\caption{Xiaomi Poco F1 channel sampling periods ($\mu$s).}
\begin{tabular}{@{}r|cccc@{}}
\toprule
\textbf{Sensor} & \textbf{$T_{c}$} & \textbf{$T_{tr}$} & \textbf{$T_{sync}$} & \textbf{$T_{end}$}  \\ \midrule
\textbf{AC} & 10000 & 7500 & 5000 & 2500  \\
\textbf{GR} & 40000 & 20000 & 10000 & 5000  \\
\textbf{GY} & 20000 & 15000 & 10000 & 5000  \\
\textbf{LA} & 40000 & 20000 & 10000 & 5000  \\
\textbf{MF} & 25000 & 20000 & 15000 & 10000  \\
\textbf{RV} & 20000 & 15000 & 10000 & 5000  \\ \bottomrule
\end{tabular}
\label{tab:pocof1_cc_periods}
\end{table}
\begin{table}[H]
\centering
\caption{Motorola Moto G5 channel sampling periods ($\mu$s).}
\begin{tabular}{@{}r|cccc@{}}
\toprule
\textbf{Sensor} & \textbf{$T_{c}$} & \textbf{$T_{tr}$} & \textbf{$T_{sync}$} & \textbf{$T_{end}$}  \\ \midrule
\textbf{AC} & 40000 & 30000 & 20000 & 10000  \\
\textbf{GY} & 30000 & 20000 & 10000 & 5000  \\\bottomrule
\end{tabular}
\label{tab:motog5_cc_periods}
\end{table}

%%
%% End of file `sample-sigplan.tex'.

\section{Profiled Applications}
\label{appendix:fingerprinted_apps}

Successfully profiled applications, taken from the top 250 Android applications~\cite{android:rank} (as of February 2022), are shown in Table~\ref{tab:app_fingerprinting_mobile}.

\begin{table*}[t]
\centering
\caption{Detected apps, category, triggered sensor(s) and sampling period(s).}
\begin{threeparttable}
\renewcommand{\arraystretch}{1.22}
\begin{tabular}{r|c|c}
\toprule
\textbf{Application} & \textbf{Category} & \textbf{Detected Sensors and Sampling Periods}  \\\midrule
\emph{AliExpress} & Shopping & \cellcolor{yellow!40}AC (SENSOR\_DELAY\_UI) \\
\emph{Among Us} & Game (Action) & \cellcolor{yellow!40} AC (SENSOR\_DELAY\_GAME) \\
\emph{Ashphalt 8} & Game (Racing) & \cellcolor{yellow!40} AC (SENSOR\_DELAY\_GAME) \\
\emph{Banana Kong} & Game (Action)& \cellcolor{yellow!40} AC (SENSOR\_DELAY\_GAME)\\
\emph{B612 Photo/Video Editor} & Photography & \cellcolor{yellow!40}AC (SENSOR\_DELAY\_UI)\\
\emph{Brain Out} & Game (Puzzle) & \cellcolor{green!40} AC, GR, GY, MF, RV (SENSOR\_DELAY\_GAME)\\
\emph{Brain Test} & Game (Trivia) & \cellcolor{yellow!40} AC (SENSOR\_DELAY\_GAME)\\
\emph{Beauty Plus} & Photography & \cellcolor{green!40} RV (SENSOR\_DELAY\_UI)\\
\emph{CamScanner} & Productivity & \cellcolor{yellow!40} AC (SENSOR\_DELAY\_UI)\\
\emph{Canva} & Art \& Design & \cellcolor{green!40}AC, GY, MF (SENSOR\_DELAY\_UI)\\
\emph{Call of Duty Mobile} & Game (Action)  & \cellcolor{green!40} AC (SENSOR\_DELAY\_GAME), GY (SENSOR\_DELAY\_NORMAL)\\
\emph{Coin Master} & Game (Casual) & \cellcolor{yellow!40}AC (SENSOR\_DELAY\_GAME)\\
\emph{Crossy Road} & Game (Action) & \cellcolor{yellow!40}AC (SENSOR\_DELAY\_GAME)\\
\emph{CSR Racing 2} & Game (Racing) & \cellcolor{yellow!40}AC (SENSOR\_DELAY\_GAME)\\
\emph{Dr.\ Driving} & Game (Racing) & \cellcolor{yellow!40}AC (SENSOR\_DELAY\_GAME)\\
\emph{eFootball PES 2021} & Game (Sports) & \cellcolor{yellow!40} AC, GY, MF (SENSOR\_DELAY\_GAME) \\
\emph{Extreme Car Driving} & Game (Racing) & \cellcolor{yellow!40} AC (SENSOR\_DELAY\_GAME) \\
\emph{Gaana Music App} & Music \& Audio & \cellcolor{yellow!40} AC (SENSOR\_DELAY\_UI)\\
\emph{Garena Free Fire} & Game (Action) & \cellcolor{yellow!40} AC (SENSOR\_DELAY\_GAME)\\
\emph{Gojek} & Travel \& Local & \cellcolor{yellow!40} AC (10ms), MF (SENSOR\_DELAY\_GAME) \\
\emph{Google Fit} & Health \& Fitness & \cellcolor{yellow!40} AC (10ms), MF (SENSOR\_DELAY\_GAME) \\
\emph{Google Maps} & Travel \& Local & \cellcolor{green!40} AC (10ms), GR, LA, MF (SENSOR\_DELAY\_GAME)\\
\emph{Grab} & Food \& Drink & \cellcolor{green!40} AC (10ms), MF (SENSOR\_DELAY\_GAME)\\
\emph{Granny} & Game (Arcade) & \cellcolor{yellow!40} AC (SENSOR\_DELAY\_GAME)\\
\emph{Helix Jump} & Game (Action) & \cellcolor{yellow!40}AC (SENSOR\_DELAY\_GAME) \\
\emph{Hungry Shark Evolution} & Game (Action) & \cellcolor{yellow!40}AC (SENSOR\_DELAY\_GAME) \\
\emph{Idle Minor Tycoon} & Game (Simulation) & \cellcolor{yellow!40}AC (SENSOR\_DELAY\_GAME)  \\
\emph{iFood Delivery de Comida} & Food \& Drink & \cellcolor{yellow!40} AC (SENSOR\_NORMAL\_UI) \\
\emph{Last Day on Earth: Survival} & Game (Action) & \cellcolor{yellow!40} AC (SENSOR\_DELAY\_GAME) \\
\emph{Ludo King} & Game (Board) & \cellcolor{yellow!40} AC (SENSOR\_DELAY\_GAME) \\
\emph{Mobile Legends} & Game (Action) & \cellcolor{green!40} AC (15ms)\\
\emph{Modern Combat 5} & Game (Action) & \cellcolor{green!40} GY (15ms)\\
\emph{Mortal Kombat} & Game (Action) & \cellcolor{yellow!40} AC, GY, MF (SENSOR\_DELAY\_GAME)\\
\emph{My Talking Tom} & Game (Casual) & \cellcolor{yellow!40}AC (SENSOR\_DELAY\_GAME)\\
\emph{Need for Speed: No Limits} & Game (Racing) & \cellcolor{yellow!40} AC (SENSOR\_DELAY\_GAME)\\
\emph{PK XD} & Game (Adventure) & \cellcolor{yellow!40}AC (SENSOR\_DELAY\_GAME)\\
\emph{Pok\'emon Go} & Game (Adventure) &\cellcolor{green!40} \cellcolor{green!40} AC (10ms), GR, GY, MF, LA, RV (SENSOR\_DELAY\_GAME)  \\
\emph{PUBG Mobile} & Game (Action) & \cellcolor{green!40}AC, GY (SENSOR\_DELAY\_GAME)  \\
\emph{Real Racing 3} & Game (Racing) & \cellcolor{yellow!40}AC (SENSOR\_DELAY\_GAME)  \\
\emph{Sberbank Online} & Finance & \cellcolor{green!40} AC (23ms) \\
\emph{Shadow Fight 2} & Game (Action) & \cellcolor{yellow!40}AC (SENSOR\_DELAY\_GAME)\\
\emph{Shadow Fight 3} & Game (Action) & \cellcolor{yellow!40}AC (SENSOR\_DELAY\_GAME)\\
\emph{Smule} & Music \& Audio & \cellcolor{yellow!40}AC (SENSOR\_DELAY\_UI)  \\
\emph{Sniper 3D} & Game (Action) & \cellcolor{yellow!40}AC (SENSOR\_DELAY\_GAME)\\
\emph{Special Forces Group 2} & Game (Action) & \cellcolor{green!40} AC, GY (SENSOR\_DELAY\_GAME), MF (SENSOR\_DELAY\_NORMAL) \\
\emph{Stand Off 2} & Game (Action) & \cellcolor{yellow!40}AC (SENSOR\_DELAY\_GAME)\\
\emph{Subway Surfers} & Game (Arcade) & \cellcolor{yellow!40}AC (SENSOR\_DELAY\_GAME) \\
\emph{Tango} & Social & \cellcolor{yellow!40}RV (SENSOR\_DELAY\_NORMAL)\\
\emph{Top Eleven Soccer Manager} & Game (Sports) & \cellcolor{yellow!40}AC (SENSOR\_DELAY\_GAME)  \\
\emph{Traffic Rider} & Game (Racing) & \cellcolor{yellow!40}AC (SENSOR\_DELAY\_GAME)\\
\emph{Trivia Crack} & Game (Trivia) & \cellcolor{yellow!40}AC (SENSOR\_DELAY\_UI)  \\
\emph{Uber} & Maps \& Navigation & \cellcolor{yellow!40}AC (10ms), MF (SENSOR\_DELAY\_GAME) \\
\emph{Uber Eats} & Food \& Drink & \cellcolor{yellow!40}RV (SENSOR\_DELAY\_NORMAL) \\
\emph{War Robots Multiplayer} & Game (Action) & \cellcolor{yellow!40}AC (SENSOR\_DELAY\_GAME)\\
\emph{Waze} & Maps \& Navigation & \cellcolor{green!40}AC, GR, LA (SENSOR\_DELAY\_GAME), MF (SENSOR\_DELAY\_UI)\\
\emph{Yandex Go} &  Maps \& Navigation & \cellcolor{green!40}AC, MF (SENSOR\_DELAY\_UI)\\
\emph{YouCam Makeup} & Photography & \cellcolor{yellow!40}AC (SENSOR\_DELAY\_UI)\\\bottomrule
\end{tabular}
\begin{tablenotes}
\item Green rows: Unique sensor/sampling combinations; Yellow: Conflicting parameters with another app. 
\end{tablenotes}
\end{threeparttable}
\label{tab:app_fingerprinting_mobile}
\end{table*}

\end{document}